\documentclass[a4paper,11pt]{article}
\pdfoutput=1 % if your are submitting a pdflatex (i.e. if you have
\usepackage{jheppub} % for details on the use of the package, please
\usepackage{physics}
\usepackage[T1]{fontenc} % if needed
\usepackage{appendix}
\usepackage{amsmath, amssymb, mathtools}
\usepackage{amsfonts}
\usepackage{stmaryrd}
\usepackage{mathtools}
\usepackage{graphicx}
\usepackage{subcaption}

\title{Semiclassics, branes, and extremality}

\author{Adolfo Holguin}

\affiliation{School of Mathematics and Hamilton Mathematics Institute, 17 Westland Row Trinity College Dublin, Dublin 2, Ireland}

\emailAdd{holguina@tcd.ie}

\abstract{We revisit the problem of computing extremal and non-extremal three point functions of semiclassical probes with single trace operators and point out certain inconsistencies in previous approaches in the literature. We clarify the roles of wavefunctions and  averaging over moduli, concluding that holographic computations may be performed with or without averaging. By carefully implementing the extrapolate dictionary for extremal correlators we explain the origin of the apparent mismatch between supergravity and CFT for extremal correlators involving giant gravitons in type IIB supergravity. We propose an ansatz for the wavefunctions of half-BPS giants which reproduces large $N$ limit of certain extremal two and three point functions in $\mathcal{N}=4$ SYM.}

\begin{document} 
\maketitle
\flushbottom

\section{Introduction}
Large $N$ conformal field theories are expected to describe theories of quantum gravity in $AdS$ spaces \cite{Maldacena:1997re}. While there has been impressive progress towards the solution of the spectral problem for light operators in integrable set-ups, there is still much to understand about how local bulk physics arises in the dual boundary description. This is particularly sharp in the context of operators dual to solitonic probes in the bulk, such as long strings and branes, where the bulk effective theory can contain within itself full-fledged quantum field theories living on the probe.  In such cases the encoding of the bulk effective theory on the boundary is far from clear, but some progress has been made in the past in understanding the spectrum of excitations of some simple probe set-ups. Of course to address more refined questions one needs to have control of the correlation functions involving heavy operators in the boundary theory.   A parallel motivation to study correlators of heavy operators is the expectation arising from semiclassical considerations that the relevant physics are controlled by the large parameters involved in the problem \cite{Hellerman:2015nra, Monin:2016jmo}. So far this idea has been made precise in scaling limits, such as the large charge 't Hooft limits \cite{Caetano:2023zwe, Brown:2025cbz}, and it is still unclear if these techniques can be extended to more physically relevant corners of parameter space such as those where the charges of the operators scale with $N$. 

An attractive set-up to test these ideas is the study of giant graviton operators in $\mathcal{N}=4$ SYM. Holographic correlation functions involving giant gravitons were considered in the past \cite{Bissi:2011dc,Caputa:2012yj,Bak:2011yy, Hirano:2012vz, Lin:2012ey, Caputa:2012dg} and certain puzzles involving mismatched of extremal correlators were raised.  More recently it was emphasized that a proper semiclassical treatment requires performing certain averages \cite{Yang:2021kot}, and agreement was found between all non-extremal three point functions of giants and single trace operators \cite{Yang:2021kot, Holguin:2022zii}. However the analysis of \cite{Yang:2021kot} failed to solve the issues with extremal correlators, although the consistency of the analytic continuation proposed by \cite{Lin:2012ey, Kristjansen:2015gpa} was argued for in \cite{ Holguin:2022zii}.  To some degree this settles the issues regarding the computation of \cite{Bissi:2011dc}, in the sense that the non-extremal correlators can be safely analytically continued to give their extremal values. This is unsatisfactory for various reasons. One issue was already raised in \cite{Yang:2021kot}, where it was noted that the extremal correlators are sensitive to the precise details of the operator basis used to compute them in the CFT. Particularly the correlation functions computed with the so-called single particle basis lead to vanishing extremal correlators while agreeing on non-extremal correlators with the more conventional Schur basis. This leads one to question whether one or another basis is the correct one, what the operator dual to giant graviton really is,  or whether these questions are even well defined. Another problem is that from the point of view of CFT kinematics extremal three point functions are qualitatively different from their non-extremal counterparts. The importance of this issue was already noted in earlier work \cite{DHoker:1999jke}, where it was realized that bulk extremal vertices vanish and the extremal correlator is due purely to boundary interactions.  We will see that the issues with holographic extremal correlators involving giants can be  traced back to a subtle error in applying the holographic dictionary, although addressing this issue does not directly allows us to compute them and more input is needed. Part of the goal of this work is to clarify these issues.

More generally the purpose of revisiting these calculations is to clarify the roles of averaging in holographic correlators. There are various proposals for semiclassical correlators involving strings, branes, and geometries \cite{Tseytlin:2003ac, Roiban:2010fe, Zarembo:2010rr, Klose:2011rm, Janik:2010gc, Skenderis:2007yb}. This however leads to a puzzle; the main lesson from \cite{Bajnok:2014sza,Yang:2021kot} is that averaging over moduli is essential for reproducing correlation functions of semiclassical states, yet this averaging procedure is missing from computations of expectation values of chiral primaries in BPS backgrounds \cite{Skenderis:2007yb, Caputa:2012dg}. The essential tension here lies in the fact that expectation values of light operators in heavy backgrounds, one-point functions, are not equivalent to OPE coefficients.  A natural question is then how to relate the one-point function computations on both sides of the duality without averaging. Since there should be a computation relating the un-averaged correlators on both sides, this points to an inconsistency of the prescription of \cite{Bajnok:2014sza,Yang:2021kot}.  What we find is that this inconsistency for BPS correlators arises from an incorrect implementation of semiclassical techniques and of the holographic dictionary. In particular the treatment of the parameter $\tau_0$ as a moduli suggested by \cite{Bajnok:2014sza} is inconsistent for BPS probes. This allows us to determine the one-point functions of chiral primaries in the background of a giant unambiguously. The resulting one-point functions agree precisely with those computed in the background of the BPS coherent state operators \cite{Berenstein:2013md, Berenstein:2022srd, Lin:2022wdr}, for instance 
\begin{equation}
    \ket{z_0}= \det \left(Z-z_0\right)\ket{0}.
\end{equation}

This paper is organized as follows. In section \ref{sec 2} we review previous calculations of heavy-heavy-light correlators of giant gravitons and gravitons in type IIB supergravity, paying close attention to the effects of boundary terms and their relation to extremal correlators. In section \ref{sec 3} we revisit these calculations by carefully implementing the extrapolate dictionary \cite{Hamilton:2006az}. In particular we find that the variational problem for maximally charged and non-maximally charged chiral primaries lead to qualitatively different calculations for the HHL correlator.  In section \ref{sec 4} we propose a supergravity computation that reproduces extremal two and three point functions of giant gravitons computed in gauge theory. Finally in \ref{discussion} we conclude and comment on future directions.

\section{HHL correlators and semiclassics}\label{sec 2}
Before discussing calculations in holography we will review some general aspects about semiclassics. Following \cite{Bajnok:2014sza}, we will place particular emphasis on boundary terms, and to the counting of zero modes. Many of these features are already present in simpler toy models so it will be useful to clarify some of those concepts there. 

Consider a free non-relativistic particle constrained to move on a circle of unit radius. The action is  
\begin{equation}
    S= \int d\tau \,L(\phi, \dot{\phi})=\frac{M}{2} \int d\tau \, \dot{\phi}^2.
\end{equation}
The equations of motion follow from the variational principle, with some boundary terms that are often ignored because the solution to the equations of motion are fixed by the boundary conditions of the problem. This would be the end of the story if were were doing classical mechanics, but in quantum mechanics we are instructed to perform a path integral to compute transition amplitudes

\begin{equation}
\bra{\phi_{\text{out}}} e^{i \int_{-T}^T \, d\tau\,H} \ket{\phi_{\text{in}}}
= \int\limits_{\mathclap{\substack{\phi(T)=\,\phi_{\text{out}} \\ \phi(-T)=\,\phi_{\text{in}}}}} D\phi \; \exp\left[i \int d\tau \, L(\phi, \dot{\phi})\right].
\end{equation}
This is the standard picture for Dirichlet boundary conditions. More precisely the boundary terms that arise from varying the action are 
\begin{equation}
\begin{aligned}
    \delta S&= -M\int d\tau \left( \delta\phi \,\ddot{\phi}\right)+ M\int d\tau \partial_\tau\left( \delta\phi \,\dot{\phi}\right)  \\
    &= -M\int d\tau \left( \delta\phi \,\ddot{\phi}\right)+ \delta\phi(T) J(T)- \delta\phi(-T) J(-T).
\end{aligned}
\end{equation}
The variational principle is well-defined only if the boundary terms vanish; this can be done by either imposing Dirichlet boundary conditions $\delta\phi(\pm T)=0$, or Neumann boundary conditions $J(\pm T)= J_\pm$, or a mixture of both. If we want to compute transition amplitudes between fixed positions we should chose Dirichlet boundary conditions, while for fixed angular momentum transition amplitudes we need to impose Neumann boundary conditions. 

But for the variational principle to be well-defined with Neumann boundary conditions one needs to supplement the action with as additional boundary term
\begin{equation}
    S'= S - \phi_{\text{out}} \,J' + \phi_{\text{in}} \,J
\end{equation}
in which case the vanishing of the variation of the action imposes the boundary conditions
\begin{equation}
    J(T)= J',\;\;\; J(-T)= J.
\end{equation}
Because now the final and initial positions are unfixed we must integrate over them ( possibly with a measure factor):
\begin{equation}
\bra{J'} e^{i \int_{-T}^T \, d\tau\,H} \ket{J}
= \int d\phi_{\text{out}} \,d\phi_{\text{in}}\;\int\limits_{\mathclap{\substack{\phi(T)=\,\phi_{\text{out}} \\ \phi(-T)=\,\phi_{\text{in}}}}} D\phi \; \exp\left[i \int d\tau \, L(\phi, \dot{\phi})- i\phi_{\text{out}} \,J' + i\phi_{\text{in}} \,J\right].
\end{equation}
This is of course what one obtains in the operator picture by inserting a complete set of states and using the fact that $J$ and $\phi$ are canonically conjugate variables. The main point here is that to perform path integral calculations with fixed momentum states one needs to introduce additional wavefunction factors that arise from the boundary terms needed for the variational principle to make sense. Now we can take the limit $J,J'\gg 1$ and work semiclassically. In this case $\phi_{\text{out}}$ and $\phi_{\text{in}}$ are not independent since they are fixed by the solution to the equations of motion.  The difference $\phi_0=\phi_{\text{in}}-\phi_{\text{out}}$ is a modulus of the classical solution and the path integral is multiplied by an overall factor
\begin{equation}
    \frac{1}{2\pi}\int d\phi_0 \;e^{i \phi_0 (J-J')}= \delta(J-J')
\end{equation}
where $\phi_0$ is the value of $\phi(t)$ at an arbitrarily chosen time, such as $t=0$. This just says that since angular momentum is conserved we cannot have classical solutions where we fix the initial and final angular momenta to be different. A more interesting scenerio happens if we include an additional operator insertion.  If we drive the system for a small time interval $[-\epsilon, \epsilon]$ via an operator insertion $\tilde{\mathcal{O}}=\exp\left[i\int d\tau V(\tau)\right]$, where $V(\tau)$ can be taken to have compact support around $\tau=0$, it is no longer true that the initial and final momentum need to be equal. Alternatively we could insert a operator $\mathcal{O}(\tau=0)$ that does not commute with the Hamiltonian or angular momentum operators. The driving potential picture is more familiar in the semiclassical picture, since we can now solve the equations of motion in the three different regions and glue them together. The solutions in the "asymptotic regions" are the same as the free equations
\begin{equation}
\begin{aligned}
    \phi(\tau)&= \phi_0+ \omega(\tau-\epsilon), \;\tau>\epsilon\\
    \phi(\tau)&= \phi_0+ \omega'(\tau+\delta), \;\tau<-\epsilon\\
\end{aligned}
\end{equation}
where $\delta$ is the relative  phase shift due to driving the system and is fixed by the equations of motion. The only free independent variable now is the position of the particle at $\tau=\epsilon$, $\phi_0$, since all the other parameters are fixed by the boundary conditions of the problem. For small enough $\epsilon, \delta$ and assuming time reversal symmetry we can replace the path integration over the $\tau>\epsilon$ region by the the square root of the path integral for $-T<\tau<T$. We can do similarly for $\tau<-\epsilon$. In the end these are essentially the normalization factors $\sqrt{\bra{J}\ket{J}\bra{J'}\ket{J'}}$. Then after dividing out by the normalizations the only thing we are left with is 
\begin{equation}
 \frac{\bra{J'} \tilde{\mathcal{O}}(0)\ket{J}}{\sqrt{\bra{J}\ket{J}\bra{J'}\ket{J'}}} \simeq  \frac{1}{2\pi}\int d\phi_0\, e^{i\phi_0(J-J')} \tilde{\mathcal{O}}(0)\bigg|_{\text{on-shell}}.
\end{equation}
The same arguments can be repeated for rather general forms of the action, since the only important point was that the driving force did not change the solution much outside a small time interval. In a more general case the statement that we can ignore the effects of finite $\epsilon, \delta$ are replaced with the assumption that the insertion of the operator $\tilde{\mathcal{O}}$ does not change the saddle point significantly, which is to say $J-J'\sim O(1)$ and not of order $|J|$.\\
The point of this exercise is the fact that we are working with fixed momentum path integrals, we need to be careful to include the correct boundary terms which lead to additional "wavefunction" contributions to correlation functions. In the semiclassical approximation the final and initial positions of the particle are not independent, since we fixed the velocities, and in general this will lead to a single moduli. This is to say that the equations of motion fix the relative positions at $\tau=\pm T$, but it is impossible to fix the absolute positions using the equations of motion. This is a generic property of classical solutions with fixed charge, since the particle is almost localized in position and in momentum space.

Before proceeding let us consider an important exception to the above prescription. In the discussion above it was assumed that the operator $\tilde{O}$ did not commute with the Hamiltonian of the system, in which case the insertion of the operator $\mathcal{O}$ is computing by its saddle point value on the classical solution, averaged over all classical solutions. One should also perform the path integral over the fluctuations around the classical solution which leads to additional ``source" vertices in the action. In string computations the source terms are vertex operators for instance.  If the operator $\tilde{O}$ happens to commute with the Hamiltonian of the system, then the path integral description for the expectation value of $\mathcal{O}(t)$ is invalid. This is because implicitly in the derivation of the path integral formalism one needs to perform a Trotterization of the Hamiltonian and insert a complete basis of states after each time step. If the operator commutes with Hamiltonian the operatator never enters the path integral but rather changes the initial or final state of the system. This point will be crucial for understanding extremal correlators.
\subsection{Giant Graviton Correlators}
The kinds of probes that we will focus on are the half-BPS sphere giant graviton solutions \cite{McGreevy:2000cw, Hashimoto:2000zp, Witten:1998xy}. the relevant parts of the action of the D3 brane are
\begin{equation}
    S= -\frac{N}{2\pi^2} \int d^4\sigma \left(\sqrt{-h}- P[C_4]\right)
\end{equation}In global $AdS_5\times S^5$ coordinates the solution is given by
\begin{equation}
\begin{aligned}
    \rho &=0\\
    \theta &= \theta_0\\
    \tau&=t\\
    \phi(t)&= t+ \phi_0\\
    \sigma_i&= \chi_i,
\end{aligned}
\end{equation}
where $\chi_i$ are the coordinates on an $S^3\subset S^5$. It can be shown that this embedding satisfies a BPS bound $\mathcal{H}= P_\phi=J$, and as a result the brane effectively behaves as a massive particle sitting at the center of global $AdS$. These solutions are dual to half-BPS sub-determinant operators in $\mathcal{N}=4$ SYM \cite{Corley:2001zk}. A calculation for the (extremal) holographic three point function of a single trace chiral primary with giant gravitons was proposed in \cite{ Bissi:2011dc} following the prescriptions of \cite{Zarembo:2010rr}. In that prescription the expectation value of a chiral primary is obtained by computing the variation of the action with respect to the supergravity modes dual to the single trace operator and substituting a bulk-to-boundary propagator for the relevant field. The dictionary between the supergravity modes and the chiral primaries is \cite{Kim:1985ez}
\begin{equation}
\begin{aligned}
    &\delta g_{\mu \nu}= 2L g_{\mu \nu}s^L(X)Y_L(\Omega_5)+\left[-\frac{16}{5}L g_{\mu \nu}+ \frac{4}{L+1} \nabla_{(\mu} \nabla_{\nu)}\right] s^L(X)Y_L(\Omega_5)\\
&\delta g_{\alpha \beta}= 2 L g_{\alpha \beta} s^L(X) Y_L(\Omega_5)\\
&\delta C_{\mu_1 \mu_2 \mu_3 \mu_4}= -4 \epsilon_{\mu_1 \mu_2 \mu_3 \mu_4 \mu_5} \nabla^{\mu_5} s^L(X) Y_L(\Omega_5)\\
&d C_{\alpha_1 \alpha_2 \alpha_3 \alpha_4} = 4 \epsilon_{\alpha\alpha_1 \alpha_2 \alpha_3 \alpha_4} s^L(X) \nabla^{\alpha} Y_L(\Omega_5),
\end{aligned}
\end{equation}
and the correlator was computed by 
\begin{equation}
    \langle O_\Delta \rangle = -\frac{\delta S}{\delta s_0},
\end{equation}
with 
\begin{equation}
    s^L= s_0 \; \mathsf{N}_L K_L,
\end{equation}
where $K_L$ is the bulk to boundary propagator ending on the brane.
In \cite{Bissi:2011dc} it was found that this prescription gives inconsistent answers for extremal correlators, although \cite{Caputa:2012yj} found perfect agreement between the gauge theory and supergravity calculations for some non-extremal correlators.  \\
To draw some analogies with the toy model of the previous section, the variation of the action with respect to the supergravity fields can be interpreted as the driving, or more precisely source term $\int d\tau V(\tau)$, while $\phi_0$ is the modulus for the spontaneously broken rotational symmetry.  The natural choice of Hamiltonian for half-BPS probes is the BPS Hamiltonian
\begin{equation}
    H_{\text{BPS}}= H- J.
\end{equation}
The chiral primaries operators considered in \cite{Caputa:2012yj} are eigenstates of this Hamiltonian
\begin{equation}
   -i [H_{\text{BPS}}, O_{(\Delta, k)}]\,= (\Delta-k) O_{(\Delta, k)}.
\end{equation}
For now let us restrict to positive charge states $k>0$ for sake of clarity.
This means that although the correlation is independent of the insertion time of light operator in the co-rotating frame, the operator cannot be moved to act only on the initial or final states for generic $k$.  The non-extremality condition then is analogous to the assumption that the operator $\tilde{O}$ does not commute with the Hamiltonian, which hints at something special happening for maximally charged operators. The computations done in \cite{Yang:2021kot, Holguin:2022zii} amount to applying the formalism of the previous section to those carried out in \cite{Bissi:2011dc}
\subsection{To average or not to average: counting moduli}
Now we will review the computation of \cite{Yang:2021kot} and clarify where we believe their implementation of the orbit average idea is erroneous. First we note that their impletation of the idea comes from adapting the techniques of \cite{Bajnok:2014sza} used for semiclassical string worldsheet calculations. In that context the claim is that a classical solution to the equations of motion coming from the Polyakov action there is actually at least a one parameter family of solutions arising from shifting the worldvolume time coordinate, schematically
\begin{equation}
    X(\tau, \sigma)\rightarrow X(\tau+ \tau_0, \sigma).
\end{equation}
The prescription for averaging over this moduli space of solutions proposed there was that the saddle point contributions to correlators should be replaced by
\begin{equation}
  \int d\sigma d\tau \hat{V}_L(\sigma, \tau)\rightarrow\lim_{T\rightarrow\infty} \frac{1}{T} \int_{-\frac{T}{2}}^{\frac{T}{2}}d\tau_0 \int d\sigma d\tau \hat{V}_L(\sigma, \tau+\tau_0).
\end{equation}
In this equation $\hat{V}_L$ is the vertex operator on the worldsheet dual to a chiral primary, which is given by inserting the relevant metric perturbations into the Polyakov action. For the D3 brane this was the variation of the action with respect to the supergravity modes. Implicitly for the string solutions considered in \cite{Janik:2011bd}, the modulus $\tau_0$ can be associated to the motion on the $S^5$. The averaging prescription was justified in analogy with the semiclassical approximation for systems without periodic motion. Additional "wavefunction" factors were presented due to approximate forms for the vertex operators associated to the classical string background. 

This proposal was adapted to  giant graviton correlators to be\footnote{Note that in the proposal \cite{Yang:2021kot} the minus sign in front of the action is missing. This minus sign appears because the calculation is implicitly done in Euclidean signature.} 
\begin{equation}
 \bra{J+k} O_\Delta \ket{J}= -\underbrace{\int \frac{d\phi_0}{2\pi}\vphantom{X}}_{\substack{\text{broken rotation}\\\text{ zero mode}}} \,\textcolor{red}{\underbrace{\int d\tau_0}_{\substack{\text{broken dilatation}\\\text{ zero mode}}}}\;\underbrace{ e^{-i k \phi_0}\, e^{-k \tau_0}}_{\text{wavefunctions}}\,\frac{\delta S(\phi_0, \tau_0)}{\delta s_0}.
\end{equation}
This proposal already differs from the \cite{Bajnok:2014sza} in that the normalization of the integration over the $\tau_0$ modulus is missing. The interpretation of $\tau_0$ is also different, since it is not to be identified with the Lorentzian time coordinate but rather to a dilatations. Although these expressions come from well motivated heuristics, it is hard to say whether this prescription is correct for  general calculations. In particular we believe that identifying $\tau_0$ with a modulus of the classical giant profile is incorrect. 

A related issue already appears in the context of the motion of a massive particle on $AdS_2$. The solution to the equations of motion is 
\begin{equation}
    (x(\tau)-x_0)^2+ z(\tau)^2= R
\end{equation}
with boundary conditions at a radial cut off $(x(-s/2)=0, z(-2/s)=\epsilon)$ and $(x(s/2)=x_0, z(s/2)=\epsilon)$. Simply shifting $\tau\rightarrow\tau+\tau_0$ would break the boundary conditions, so $\tau_0$ cannot be taken to be a true moduli for Dirichlet boundary conditions.

In the sphere giant graviton calculation the situation is more clear because we can work in a static gauge where the worldvolume time coordinate is the global time coordinate $\sigma^0= t$, and the spacelike directions wrap an $S^3$ inside of the $S^5$ which reduces to the problem to an effective one dimensional problem. The simplest solutions to the equations of motion are those that lead to the BPS condition
\begin{equation}
    H-J=0.
\end{equation}
Generically these break rotational symmetry along an axis of the $S^5$ and time translation symmetry. However because of the BPS condition the generator $H_{\text{BPS}}= H-J$ remains a symmetry of the problem. This means that although one might naively expect that the target space coordinates on the worldvolume have two independent moduli,
\begin{equation}
\begin{aligned}
      \tau(\sigma)&\rightarrow\tau(\sigma) + \tau_0\\
       \phi(\sigma)&\rightarrow\phi(\sigma) + \phi_0
\end{aligned}
\end{equation}
in reality they cannot both be independent of each other. This is to say that the prescription described in \cite{Bajnok:2014sza} needs to be implemented with care, and that one should properly count the number of broken symmetries. On the other hand $\tau_0$ was interpreted as a dilatation modulus in \cite{Yang:2021kot} which seems to invalidate the objection above. This was justified there by stating that the bulk calculation could be restricted to a spacelike time slice $\tau=0$ in global AdS, and without this moduli integration the calculation cannot give the correct answer. We will see later that the interpretation of $\tau_0$ as a ``dilation modulus" can be made sense of as a sort of Euclidean state preparation, but the calculation as described in \cite{Yang:2021kot} is not correct as it does not follow from the extrapolate dictionary for AdS/CFT.

In the bulk description the correct calculation is to first solve the equations of motion in the presence of the brane, and then compute the effective action by expanding around the solution in field variations. There are two kinds of perturbations, those restricted to the brane which are dual to modifications of the determinant operator which we will call $\delta X$, and perturbations of the bulk supergravity fields $\delta g$. In general these are coupled by the effective action of the brane 
\begin{equation}
    S_{D3}= \int d\tau d\Omega_3 \left[ \mathcal{L}+ \frac{\delta\mathcal{L}}{\delta  g}\delta g+ \frac{\delta\mathcal{L}}{\delta  X}\delta X+ \frac{\delta^2\mathcal{L}}{\delta  g \delta X}\delta g \delta X + \dots\right]+ \texttt{boundary terms}
\end{equation}
The first order variations with respect to the worldvolume fields vanish on-shell due to the equations of motion of the brane, but the first order variations with respect to the bulk fields do not need to vanish. We should also be careful to not ignore the effects of boundary terms that arise from integration by parts. If the Lagrangian is analytic in the fields then the brane-bulk interactions should begin at cubic order. The insertion of a single trace operator is dual to an insertion of a supergravity mode $s_I$ at the boundary, and the brane serves as a bulk source for these modes via the $\delta g$ couplings. very schematically the three point function comes from terms of the form
\begin{equation}
   \lim_{R\rightarrow \infty} \int d^{10} x' \sqrt{g}\;\bigg\langle 
  s_I(\rho= R,\tau=0) \delta g(x') \bigg\rangle_{\text{SUGRA}} \frac{\delta \mathcal{L}}{\delta g}\;\bold{\delta_{B}}(x')
\end{equation}
where the $\delta_{\bold{B}}$ function localized the integration over the interaction vertex to the brane. At leading order in $1/N$ the supergravity correlator is given by a differential operator (coming from the couplings in the effective of the brane) acting on a bulk-to-boundary propagator and a spherical harmonic on the $S^5$. Clearly in this calculation we must integrate over the time coordinate at which the propagator intersects the brane, which contradicts the prescription of \cite{Jiang:2023cdm} in a Lorentzian setting.

\subsection{Why computing without averaging gives the right answer}
One last issue to address  is why the method of \cite{Zarembo:2010rr, Bissi:2011dc} seems to give correct answers in some non-extremal cases \cite{Caputa:2012yj} but not in extremal ones. We should clarify what we mean here. In \cite{Yang:2021kot} it was noted that some non-extremal correlators were correctly reproduced without performing an integration over the moduli of the semi-classical solutions; the correlators in question were computed by \cite{Caputa:2012yj}. Later \cite{Yang:2021kot} demonstrated that this prescription gives wrong answers if one uses generic polarizations for the spherical harmonics. This is not strictly correct, and as we will see the reason that the calculations of \cite{Caputa:2012yj} succeed is because there is an implicit averaging in their calculations. 

Now I will review some aspects of the computations of \cite{Caputa:2012yj}, but in a language that makes the implicit "averaging" manifest. This will also clarify why computations of holographic one point functions in backreacted LLM backgrounds do not seem to require any averaging, even though the states are semiclassical and they break the same symmetries as the giant graviton. The first point is that any given half BPS solution to the equations of motion to the D3 brane effective action will break the $R$-symmetry spontaneously, as long as the brane is not a maximal sphere giant graviton. This is just restating the fact that in the fixed angular momentum picture we had a moduli $\phi_0$. Now we will work with the path integral with Dirichlet boundary conditions instead and deal with the broken symmetries correctly. This means that without loss of generality we can set $\phi_0=0$.

In the presence of the brane, the isometries of the $AdS_5\times S^5$ background are broken to $\mathbb{R}\times SO(4)\times SO(4)$, where $\mathbb{R}$ is generated by $H_{\text{BPS}}$. A chiral primary with generic $R$-symmetry polarization is not covariant with respect to this residual symmetry. If we want to compute quantities that can be compared on both sides of the duality we better use operators that transform properly under this symmetry. The only chiral primaries that are allowed to have non-zero expectation values are those which are $SO(4)_R$ invariant, and they are labeled by their $SO(2)$ R-charge $k$. These are the operators considered in \cite{Caputa:2012yj}. One nice property of working with these operators is that the integration over the space-like worldvolume directions of the brane trivializes. This is simply a consequence of the fact that at every step of the calculation we are preserving $SO(4)_R$ symmetry. Because the operators have fixed $R$-charge the integration over the moduli $\phi_0$ would have been trivial as well, giving a delta function. 

This can be stated in a different way. In the fixed momentum prescription we would have had to integrate over the moduli $\phi_0$ with a wavefunction factor which measures the $SO(2)$ charge difference between the initial and final states. Then one would have had to integrate over the worldvolume directions, which imposes $SO(4)_R$ symmetry. So in the end the result of the orbit average is to project the polarization of the single trace chiral primary to a particular $SO(4)_R$ invariant chiral primary. In that case we fixed the final and initial states to differ by a fixed amount of $SO(2)_R$ charge, and the proper semiclassical calculation with the averaging projects the third operator into the only possible charged state that could have a non-correlation function with the given initial and final states. But we could have very well equivalently "pushed" the wavefunction factors to the third operator from the start and worked with initial and final states with indefinite R-charge. These are equivalent computations in the end, and the "incorrect" calculations presented  by \cite{Yang:2021kot} in fact compute something entirely different. By choosing an arbitrary polarization they in fact computed a sum of correlators weighted by their charge, as opposed to a diagonal structure constant. To see this we take the appropriate spherical harmonic calculation for the choice
\begin{equation*}
   \mathcal{O}_L= \tr\left(\frac{Z+ \Bar{Z}+ Y- \Bar{Y}}{2}\right)^L \rightarrow 
\bigl(
\sin\theta\cos\phi
+ i\,\cos\theta\cos\chi_1\sin\chi_3
\bigr)^{L}
\end{equation*}
which is 
\begin{equation}\label{fakeOPE}
\begin{aligned}
\langle\mathcal{O}_L(\theta_0, \phi_0) \rangle_{\texttt{fake}}=&-\frac{N}{2\pi^2}\int\,d^4\sigma\, \cos^4 \theta_0 \; 4L\, \mathsf{N}_L \left[\left(1 - \frac{1}{L}\tan\theta_0 \partial_\theta\right)Y_L K_L - \frac{1}{2}\sec^2 \theta_0 Y_L\,K_{L+2}(t) \right]
\end{aligned}
\end{equation}
where $K_L(t)$ is the bulk-to-boundary propagator for a bulk insertion at the center of global AdS.
The integration over the $S^3$ defines $SO(4)_R$ invariant harmonics which we can extract by expanding in powers of $e^{i\phi}$. These are given by the following formula
\begin{equation}
\begin{aligned}
    \tilde{Y}_{(L,k)}(\theta, \phi)&= e^{ik\phi }\,2^{-L} \binom{L}{\frac{L-k}{2}} \sin ^L\theta  \, _2F_1\left(-\frac{1}{2}
   (L+k),-\frac{1}{2}(L-k);2;-\cot ^2\theta \right)\\
  &=\frac{(-1)^{\frac{L-|k|}{2}}}{2^L} \left(\frac{1+(-1)^{L-{|k|}}}{2}\right) \frac{1}{\frac{L-|k|}{2}+1} \binom{L}{\frac{L-k}{2}} \;\sin^k \theta\, P_{\frac{L-|k|}{2}}^{(|k|,1)}(\cos 2\theta)\, e^{i k \phi}
\end{aligned}
\end{equation}
where $P_n^{(\alpha, \beta)}(x)$ are Jacobi polynomials. These are the only components of $Y_L$ that give non vanishing contributions to the correlator. We can easily check that without averaging the answer almost matches the gauge theory calculation with a coherent state. On the gauge theory side the correlator without averaging (up to unimportant kinematic factors) is
\begin{equation}
   \bra{z_0} \mathcal{O}_L \ket{z_0}= -\frac{2}{\sqrt{L}} T_L(\sin \theta_0 \cos\phi_0)
\end{equation}
where the state is created by the shifted determinant
\begin{equation}
    \ket{z_0}= \det(Z-z_0)\ket{\Omega}
\end{equation}
and $z_0= \sin \theta_0 \cos \phi_0$ and $T_n(x)$ is a Chebyshev polynomial. Note that there is no selection rule on $L$ due to contributions from off-diagonal structure constants. This answer can be obtained by resumming the diagonal and off-diagonal structure constants with appropriate phases 
\begin{equation}
    \sum_{k=-L}^L C_{\mathcal{D}_{\Delta+k} \mathcal{D}_{\Delta} \mathcal{O}_L} e^{i k \phi_0}= -\frac{2}{\sqrt{L}} T_L(\sin \theta_0 \cos \phi_0).
\end{equation}
The diagonal and non-extremal correlators were computed holographically in \cite{Yang:2021kot} and agree with the gauge theory computations.  So term by term the result of \eqref{fakeOPE} will agree with the coherent state correlation function except for the  extremal contributions which are given by the computations of \cite{Bissi:2011dc}. One interpretation of this is that the calculation without the averaging is almost correct, and that the prescription for the extremal correlators is incorrect. Clearly issue causing the mismatch should then be something independent of the averaging procedure, and rather an inconsistency in the prescrition of \cite{Zarembo:2010rr, Kristjansen:2023ysz} when applied to extremal correlators. Other possibilities were raised in \cite{Caputa:2012yj, Yang:2021kot}. One possibility is that the sphere giant graviton is described a different kind of operator on the gauge theory, because the value of extremal correlators are basis dependent even in the large $N$ limit. Because there is no canonical basis of operators one might also conclude that extremal correlators cannot be captured by supergravity. We will comment on this possiblility later. We find these implausible given the fact that extremal vertices can be computed, although their computation is qualitatively different from non-extremal vertices \cite{DHoker:1999jke}. 

\subsection{Why averaging and not averaging both fail }
To see what is going wrong with the calculation let us concentrate on the extremal contributions to the correlators for the sphere giant graviton. According to the prescription of \cite{Zarembo:2010rr, Bissi:2011dc} the correlator is a sum of the contributions of the variations of the area and Chern-Simons terms. The contribution from the area term can be split into two kinds of terms, those which are proportional to a product of the boundary source $s_L$ and the induced metric without derivatives, and those which include derivatives. The terms without derivatives are given by 
\begin{equation}
    \delta S_{DBI}^{(1)}= -\frac{N}{2\pi^2} \int d^4 \sigma \sqrt{-h} \, 4L\, Y_L s^L(t)
\end{equation}
while those with derivatives are 
\begin{equation}
     \delta S_{DBI}^{(2)}= -\frac{N}{2\pi^2} \int d^4 \sigma \sqrt{-h} \, 2L \, Y_L h^{tt}\left[\frac{1}{L(L+1)} \partial_t^2 + \frac{L}{L+1}\right] s^L(t).
\end{equation}
The CS term contributes
\begin{equation}
 \delta S_{CS}= \frac{N}{2\pi^2}\int d^4\sigma \cos^3\theta \sin \theta \,4\,\partial_\theta Y_L s^L(t).
\end{equation}
Now consider the extremal contribution to the correlation which is captured by the spherical harmonic
\begin{equation}
    Y_{(L,\pm L)}\propto e^{\pm i L \phi} \sin^L \theta_0.
\end{equation}
In that case $\delta S_{DBI}^{(1)}$ exactly cancels with $\delta S_{CS}$ which is a consequence of the fact that the brane embedding is supersymmetric. This makes sense because the fields $s_{(L, \pm L)}$ preserve the same supersymmetry as the brane solution. This means that the incorrect calculation is only due to $\delta S_{DBI}^{(2)}$. One way of interpreting this terms is as a bulk source for the supergravity field $s_L$ with a differential operator acting on the field. For this reason it makes more sense to integrate by parts and have the derivatives act on the spherical harmonics pulled back to the brane as opposed to the fields $s_I$. On the brane the spherical harmonics become time dependent so we get
\begin{equation}
\begin{aligned}
   \delta S_{DBI}^{(2)}&= -\frac{2N}{L+1} \cos^2 \theta_0 \int dt\,  s^L\left[\partial_t^2 + L^2\right] Y_{(L,L)}(t)\\
   &-\frac{2N}{L+1} \cos^2 \theta_0 \int dt\,\partial_t \bigg( \partial_t s_L Y_{(L,L)}- s_L\partial_t  Y_{(L,L)}\bigg)\\
    &=\frac{N}{2\pi^2} \int dt\,d^3 \sigma \sqrt{-h}\, 2L\partial_t \bigg( (\partial_t-\partial_\phi) s_L Y_{(L,L)}\bigg)
\end{aligned}
\end{equation}
The first term vanishes on-shell since $\partial_t^2 Y_{(L,L)}= -L^2 Y_{(L,L)}$ and the remaining term comes purely from $t=\pm \infty$. The issue now is that we are left with a boundary term, and in the computation of the one-point function one is instructed to solve the equations of motion of the field $s_{L}$. This is problematic since boundary terms lead to modifications of boundary conditions once one imposes the well-posedness of the variational problem for the effective action. What this says is that whatever calculation captures the extremal contributions will not be a result of a bulk Witten diagram calculation, and so the form of the effective action of the brane is irrelevant and so one needs additional input relative to the non-extremal correlators. 

\section{Semiclassics and D-branes}\label{sec 3}
\subsection{Coherent State Path Integrals}
To see that we are indeed missing something in the holographic calculation let us turn to a simpler puzzle. From what we saw previously it is natural to conjecture that a classical giant graviton solution is dual to a coherent state operator of the gauge theory in the large $N$ limit. This would be consistent except that the contribution from extremal correlators. To see that the issue is rooted at the extremality of the correlator, let us consider the simplest extremal correlator; the norm of the coherent state. In the gauge theory the norm of shifted determinant is given by
\begin{equation}
    \bra{z_0}\ket{z_0}= \frac{N!}{N^N}\, e^{N |z_0|^2}=\sqrt{2\pi N} \times e^{-N \cos^2\theta_0}.
\end{equation}
Can one reproduce this in the gravity description?
A first guess would be that this computes the exponential of the action of the brane on-shell, but this is not true because the action vanishes on-shell due to supersymmetry. This mismatch is often justified with the fact that the normalization of two point functions is unphysical, in the sense that it will cancel in higher point correlators, but if the two states are truly equivalent having these normalizations match in gravity and field theory is not out of the question. \footnote{For a similar issue in the case of surface operators see \cite{IzquierdoGarcia:2025jyb}.} Another guess is that the norm computes the area of the brane, but this is also incorrect because the area has units of action (notice we set the curvature scale to one). A quantity that has the correct scaling is the Hamiltonian. For instance on the BPS solutions we have
\begin{equation}
    H= N \cos^2 \theta_0.
\end{equation}
But it is not at all clear how the Hamiltonian could appear as any kind of saddle point value, given the brane is four-dimensional and it is more natural to have the area element $\frac{N}{2\pi^2}\int d^4\sigma \sqrt{-h}\sim\cos^4\theta_0$. In any case it seems that if we want to reproduce a coherent state calculation using holography we should not try to work with the position space path integral of the giant, but rather a phase space path integral. 

To see why this is consider the time evolution of a coherent state in a one dimensional quantum mechanical system. The transition amplitude can be expressed as a path integral in the usual way
\begin{equation}
\begin{aligned}
   \bra{\alpha_f}e^{-i T H} \ket{\alpha_i}&\sim \int_{\alpha(0)= \alpha_i}^{\alpha(T)}\mathcal{D}\alpha(\tau) \mathcal{D}\alpha^*(\tau)\; \exp\left(i\int d\tau \left[K(\alpha, \alpha^*)-H(\alpha, \alpha^*)\right]\right)\\
   K(\alpha, \alpha^*)&= \frac{1}{2i}\left(\alpha\partial_\tau\alpha^*- \alpha^* \partial_\tau \alpha\right)
\end{aligned}
\end{equation}
One should be careful to not put an equality sign here because in the derivation of the path integral one encounters boundary terms that come from careful treatment of the Trotter procedure. Another way of saying this is that if we set $H=0$ we should reproduce the coherent state norm but the right hand side of the equation is ambiguously defined. The correct expression is 
\begin{equation}
    \bra{\alpha_f}e^{-i T H} \ket{\alpha_i}= \int_{\alpha(0)= \alpha_i}^{\alpha(T)= \alpha_f}\mathcal{D}\alpha(\tau) \mathcal{D}\alpha^*(\tau)\; e^{i\int_0^T d\tau \left[K(\alpha, \alpha^*)- H(\alpha, \alpha^*)\right]}\; e^{\frac{1}{2}\left(|\alpha_i|^2+ |\alpha_f|^2\right)},
\end{equation}
and the wavefunction factors $e^{\frac{1}{2}\left(|\alpha_i|^2+ |\alpha_f|^2\right)}$ come from carefully treating the boundary terms of action. 

Similarly the path integral for a sphere giant restricted to the half-BPS sector is \cite{Mandal:2005wv}
\begin{equation}
\begin{aligned}
  \mathcal{Z}_{BPS} &=\int \mathcal{D}[r^2(t)]\mathcal{D}[P_r(t)]\mathcal{D}[\phi(t)]\mathcal{D}[P_{\phi}(t)]\;\delta(r^2-P_\phi)\,\delta(P_r) \; e^{i \int dt \left[\dot{\phi}\, p_{\phi}- H\right]}\\
  &=\int \mathcal{D}[r^2(t)] \mathcal{D}[\phi(t)] \; e^{i \int dt \left[\dot{\phi}\, r^2- H\right]}=\int \mathcal{D}[\bar{z}(t)] \mathcal{D}[z(t)] \; e^{i \int dt \left[K(z,\bar{z})- H\right]} .
\end{aligned}
\end{equation}
The delta functions impose the BPS conditions as constraints on the phase space and in the last line we wrote the action in a complex polarization. This suggests that the two point function should correspond to the following path integral 
\begin{equation}
    \bra{\mathcal{B}_f} e^{-i \,T\,\tilde{\mathcal{H}}}\ket{\mathcal{B}_i}= \int_{z(0)}^{z(T)} \mathcal{D}[\bar{z}(t)] \mathcal{D}[z(t)] \; e^{i \int dt \left[K(z,\bar{z})- \tilde{\mathcal{H}}\right]} \;|\psi_{\mathcal{B}_i}|\times|\psi_{\mathcal{B}_f}|.
\end{equation}
The action in the exponential is just the effective action of a $D3$ brane written in canonical form. Without any further operator insertions the Hamiltonian acts as a phase factor which reflects the fact that the on-shell action vanishes, meaning that the overall normalization is given by the norm of the wavefunction of the brane. The same is true for extremal correlators, since they carry the same charges as the background state, so their contribution cannot come from varying the action in any meaningful way. We will see this more precisely later on. We would hope to show that the norm of the wavefunctions of the branes computed holographically agrees with the wavefunctions computed on the gauge theory. In general we cannot guarantee that this will be true, as the wavefunctions generically get renormalized from weak to strong coupling. Since the states are supersymmetric one can hope that there is a supersymmetric renormalization scheme, and then the wavefunctions should agree since the half-BPS operators are not renormalized. 

It is not surprising then that determining the expectation values of maximally charged chiral primaries in the background of the brane is equivalent to determining information about the wavefuction of the branes. 
\subsection{Toy Effective Action}
To illustrate what is happening in the case of a probe giant graviton, let us consider the simpler case of a complex massive scalar field coupled to a point-like probe in global AdS. To analyze the dynamics of this system we can work in the saddle point approximation, treating the background metric and probe configuration as fixed on a real time interval $[-T,T]$. For simplicity we can assume that the equations of motion of the probe pin it to $\rho=0$. In the global AdS quantization we would want to take $T\rightarrow \infty$ but one has to be careful about boundary terms that appear from preparing the asymptotic states of the system. This state preparation may be done by gluing a Euclidean cap at both ends of the Lorentzian cylinder and matching the boundary conditions on the gluing surfaces \cite{Skenderis:2008dg}. On each these Euclidean segments the probe will move in a geodesic in Euclidean AdS which ends at the origin of the boundary of a Poincare patch, or equivalently along a geomesic of one half of a global Euclidean $AdS$.
\begin{figure}[t]
    \centering
   \includegraphics[width=.4\textwidth]{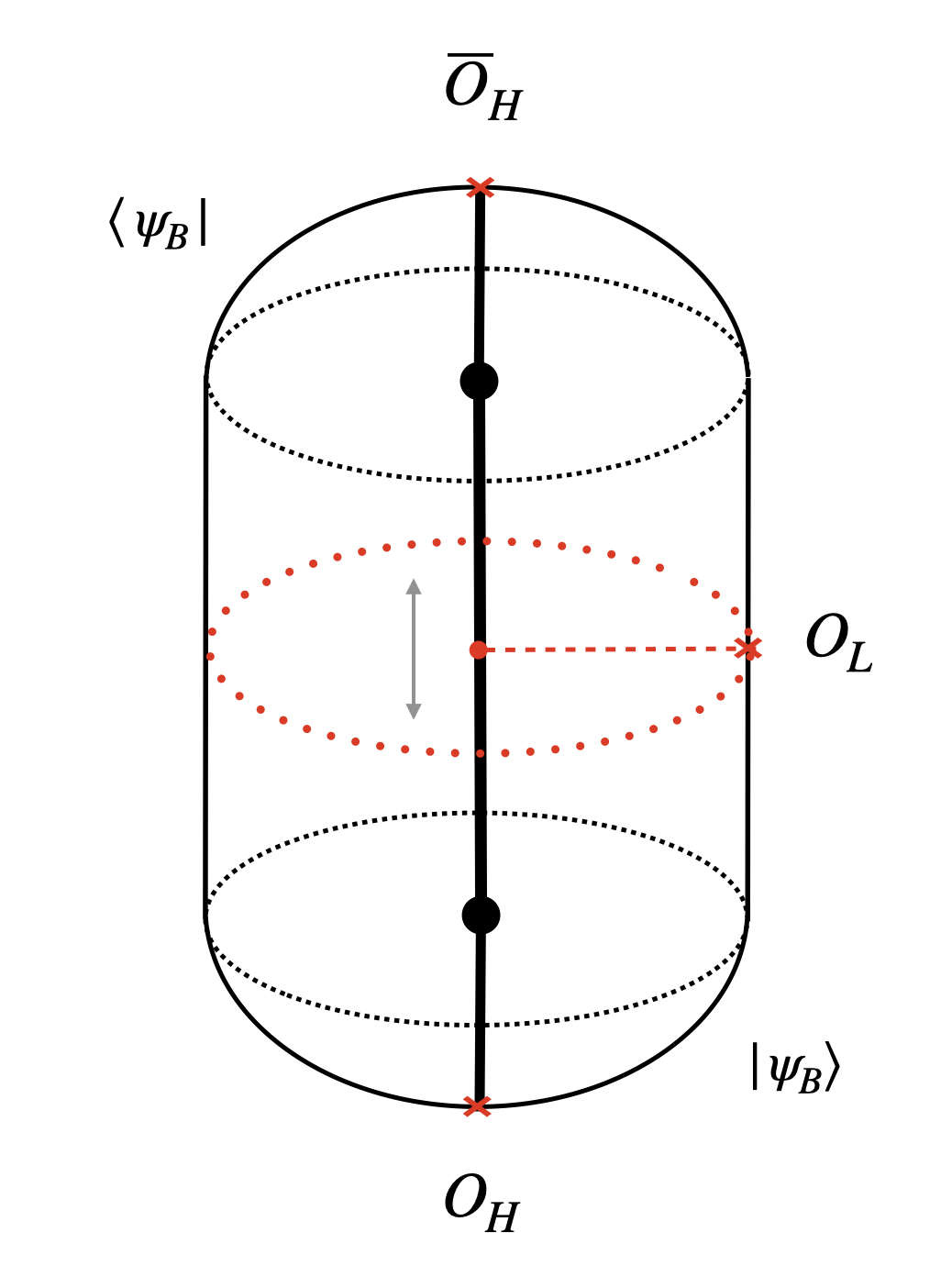}
\caption{Sketch of the Lorentzian computation of the one point function. The red dot indicated the location of the bulk vertex which is integrated along the worldvolume of the probe. The initial and final state are inserted in the Euclidean segments of the geometry. The Euclidean caps prepate the initial and final states for the Lorentzian time evolution.}
    \label{fig:placeholder}
\end{figure}

In the Lorentzian segment the effective action to the lowest order in the fields is of the form
\begin{equation}
S_{eff}= \int d^5x \sqrt{-g}\left[D_\mu \phi^* D^\mu \phi+ M^2 |\phi|^2+ \mathsf{g}\, \mathcal{D} \phi^*\,L(t) \,\delta(\rho) +c.c.\right]+ S_{\text{probe}}+ \dots
\end{equation}
where $\mathcal{D}$ is some differential operator and $L(t)$ encodes the time dependence of the probe solution. For example if $\phi$ has charge $k$, the coupling would be of the form $L(t)\sim e^{ikt}$. Because the probe sits at $\rho=0$ only the rotationally invariant parts of $\phi$ will couple to the probe. The interactions can be of the schematic form
\begin{equation}
    \mathcal{D}\phi= \alpha_{(0,0)} \phi + \alpha_{(1,0)}n^t\nabla_t \phi +\alpha_{(0,1)}n^\rho\nabla_\rho\phi+ \dots
\end{equation}
Say that $\phi$ is dual to some operator $\mathcal{O}$ on the boundary. Because the coupling to the probe acts as a source term for the path integral the bulk field $\phi$ has a non-zero one point function which can be pushed to the boundary to extract the expectation value of $\mathcal{O}$.

We also have the actions $    I= I_-+ I_+$ at the caps which are the Euclidean continuation of the above. These cap factors are related to the in and out wavefunctions as $T\rightarrow \infty$. To compute the one-point function of $\mathcal{O}$ we need to quantize $\phi$ with prescribed boundary conditions on the Euclidean segments and on the gluing surfaces. Semiclassically one needs to make sure that the variational problem for $\phi$ is well-defined. In practice this generates boundary terms that come from integration by parts and these terms must be eliminated by adding appropriate counter terms. For the action of the scalar field this is well understood so we will focus on the terms that couple the field to the probe. \\
The naive prescription for computing the one point function of $\phi$ would be to take the effective action on the Lorentzian segment and expand out to leading order in $\phi$ which would give 
\begin{equation}\label{one-point-unren}
    \langle\mathcal{O}(0)\rangle\propto \mathsf{g} \lim_{r\rightarrow\infty}\,r^\Delta\int dt \,S(t)\langle \phi(0, r)\mathcal{D}\phi(t, 0)\rangle.
\end{equation}
This is usually correct, but in the cases of extremal correlators we need to be more careful. What one really should do is to take the effective action and integrate by parts to isolate the equations of motion. For the probe coupling this is done by rewritting the bulk couplings such that they involve no derivatives 
\begin{equation}
    S_{p\phi^*}=\mathsf{g}\int d^5x \sqrt{-g}\left[\, \mathcal{D} \phi^*\,L(t) \,\delta(\rho) \right]=\mathsf{g}\int d^5x \sqrt{-g}\left[\, \phi^*\,\tilde{\mathcal{D}} L(t) \,\delta(\rho) \right]+\text{boundary terms}
\end{equation}
This would be equivalent to integration by parts of \ref{one-point-unren} except that there can be cases where the bulk coupling vanishes $\tilde{\mathcal{D}} L(t)=0$. In such cases the naive correlator comes purely from the boundary terms at $t=\pm T$. If one adds back the contributions from the Euclidean segments one would obtain similar boundary contributions on the gluing surface but the matching conditions for the fields on the gluing surface forces all the boundary contributions at $t=\pm T$ to cancel identically, leaving only boundary terms at the Poincare boundaries of the Euclidean caps. In general these remaining contributions are divergent due to usual volume divergences and they must be renormalized. After taking $T\rightarrow\infty$ we would find that these correspond to renormalization of the wavefunction of the probe, and as such they are scheme dependent. In other words the precise details of such boundary contributions encode details about the UV structure of the wavefunction of the state in global AdS. Equivalently we could have taken $T\rightarrow 0$ which would reproduce the prescription of Komatsu but without wavefunction renormalizations. In that case the "moduli" integral over $\tau_0$ in reality corresponds to the Euclidean segments defining the initial and final wavefunctions. 
 \begin{figure}
     \centering
  \includegraphics[width=.45\textwidth]{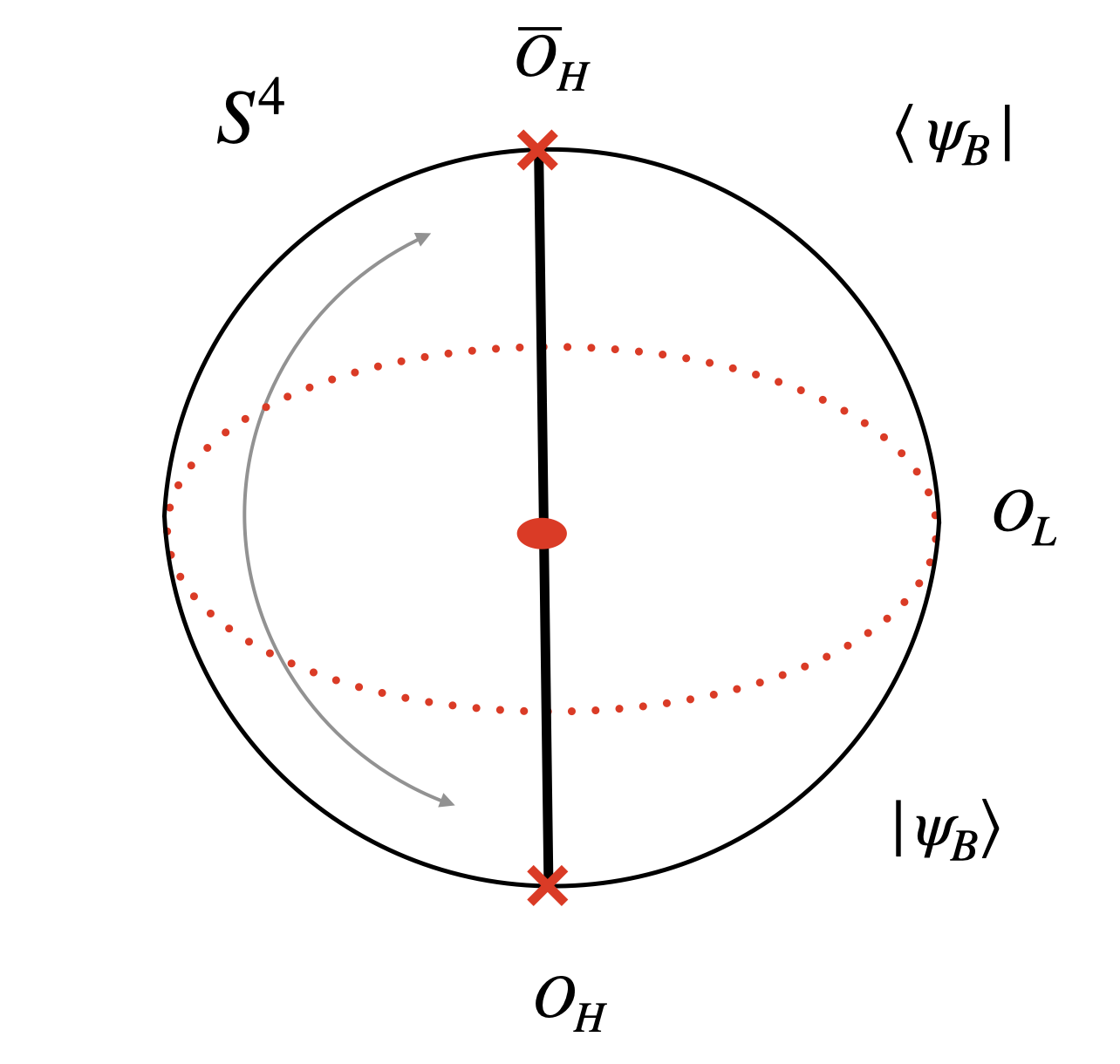}
 \caption{The Euclidean calculation implicitly performed in \cite{Yang:2021kot}. The Euclidean solution shares the $t=0$ slice with the Lorentzian one. Note the integration over the bulk vertex is completely within the Euclidean region. This is interpreted there as the integration over the moduli ``$\tau_0$''.  }
     \label{fig:euclideanfigure}
 \end{figure}
\subsection{Giant effective action revisited}
For the giant graviton the discussion applies since the couplings of the KK modes of the graviton to the brane effective action generically involve derivatives acting on the supergravity fields. To make the discussion as transparent it is useful to reduce the action of the giant to an effective one dimensional one. Concentrating on the sphere giant gravitons, the integration over the worldvolume $S^3$ can be trivialized by expanding the spherical harmonics into $SO(4)$ singlet and non-singlet parts, of which only the singlets contribute to the one-point function. The 5d field operators will have a mode expansion as
\begin{equation}
    s_L=\sum_{k=-L/2}^{L/2} \; \tilde{Y}_{(L,k)}(\theta, \phi) \; \sum_{n=0}^\infty \; \left(\mathsf{s}_{n,k} \,e^{i (\Delta+2n)t}+ 
\mathsf{s}_{n,k}^\dagger \,e^{-i (\Delta+2n)t} \right) R_{n,0}(r) +\text{non-singlets}.
\end{equation}
Here the $\tilde{Y}_{(L,k)}$ are $SO(4)$ invariant $S^5$ harmonics, and $R_{n,\ell}(r)$ are radial modes for the Klein-Gordon operator in $AdS_5$, for more details see appendix \ref{propagators}. To avoid clutter we leave the dependence on the conformal dimension of the mode $\Delta=L$ implicit. Because the sphere giants sit at the center of AdS only the lowest angular momentum mode contributes to the correlator. The modes $\mathsf{s}_{n,k}$ satisfy canonical commutation relations with the normalization 
\begin{equation}
    \bra{\Omega}\mathsf{s}^\dagger_{n,k}\mathsf{s}_{n,k}\ket{\Omega}= \mathsf{N}_L= \frac{1}{N}\frac{L+1}{4\sqrt{L}}.
\end{equation}
The coupling of $s_L$ to the brane can be decomposed into three types of contributions, those which are proportional to the target space metric (trace terms), from the coupling to the Chern-Simons term, and from a descendant of $s_I$. The trace and CS couplings combine into a single differential operator that acts only on the $S^5$ harmonic, and the descendant term gives a differential operator acting on the 5d fields themselves. The effective action takes the form
\begin{equation}
\begin{aligned}
   S_{ggs}
    &=  -N\cos^4 \theta_0\int\,dt\, \sum_{L,k} \; 4L\, e^{ikt}\left[D_\theta + h^{tt}D_{tt}\right]s_{(L,k)}\,Y_{(L,k)}(\theta_0, \phi_0) \\
    D^{(L)}_{tt}&=\frac{1}{L(L+1)} \left[\partial_t^2 + L^2\right], \;\;\;\;
    D^{(L)}_{\theta}=1-\frac{1}{L} \tan\theta_0 \partial_{\theta_0}.
\end{aligned}
\end{equation}
The differential operator $D_{tt}$ satisfies the following property when acting on the bulk-to-boundary propagator 
\begin{equation}
\begin{aligned}
      D_{tt} K_{\Delta}= K_{\Delta+2}\\
\end{aligned}
\end{equation}
which is clear since that terms comes from coupling to a descendant field. After integrating by parts we obtain the action
\begin{equation}
\begin{aligned}
   S_{ggs} 
    &=  -N\cos^4 \theta_0\int\,dt\, \sum_{L,k} \; 4L\,s_{L,k} e^{ikt}\left[D_\theta + \sec^2\theta_0\frac{(L-k)(L+k)}{L(L+1)}\right]\,Y_{(L,k)}(\theta_0, \phi_0) \\
&-N\cos^4 \theta_0\int\,dt\,\sum_{L,k} \frac{2\sec^2\theta_0}{L+1} \partial_t\left[ e^{ikt}\partial_t s_{L,k} -s_{(L,k)}\partial_t\,e^{ikt}\right]Y_{(L,k)}(\theta_0, \phi_0).
\end{aligned}
\end{equation}
The boundary of the $t$ integration lies in the Euclidean region where $\tau=-it $ is large and real. For non-extremal bulk operators $L\neq k$ the boundary terms decay fast enough that their contribution vanishes as the regulator is taken to infinity. This means that the variational problem for the non-maximally charged fields is well defined.  For maximally charged operators the bulk action vanishes identically leaving only the boundary term which in this case give a non-vanishing contribution as we take the regulator to infinity. Since the boundary terms do not vanish the variational problem for the fields $s_{(\Delta, \pm |\Delta|)}$ is not well-defined.
\subsection{Non-extremal Correlators}
For the non-maximally charged operators the resulting bulk vertex is 
\begin{equation}
     S_{ggs}^{(L,k)} 
    =  -N\cos^4 \theta_0\int\,dt\,  \; 4L\,s_{L,k} e^{ikt}\left[D_\theta - \frac{1}{2}\sec^2\theta_0\frac{(L-k)(L+k)}{L(L+1)}\right]\,Y_{(L,k)}(\theta_0, \phi_0).
\end{equation}
The computation of the one-point function of $s_{L,k}$ can be done in two equivalent ways, either by quantizing the field or by solving the equations of motion with a boundary source. To quantize the field we decompose $s_{L,k}$ into modes
\begin{equation}
\begin{aligned}
    s_{L,k}(t,r)&= \frac{1}{\sqrt{\Delta-1}}\sum_{n=0}\left(\mathsf{s}_{L,k,n}^\dagger e^{i (\Delta+2n)t}+ \mathsf{s}_{L,k,n} e^{-i (\Delta+2n)t} \right) R_n(r)
\end{aligned}
\end{equation}
where we only kept the S-wave modes on the $S^3$ inside of $AdS$. The correlator needed for the vertex gives resums into the bulk-to-boundary propagator
\begin{equation}
  \lim_{r\rightarrow \infty}    \big\langle  r^{L}  s_{L,k}(t'=0,r)s_{L,k}(t,r=0)\big\rangle= \mathsf{N}_L \,K_{L}(t,r=0; t'=0).
\end{equation}
The sum in the propagator can be expressed as a Mellin-Barnes integral
\begin{equation}
  K_\Delta(t, r=0; t'=0)= \frac{2^{(L-1)}}{2\pi i}\int_{\mathcal{I}} ds   \; \frac{\Gamma\left(\frac{\Delta}{2}-s\right) \Gamma\left(\frac{\Delta}{2}+s\right) }{ \Gamma(\Delta)} e^{-2it s},
\end{equation}
where the contour is chosen to divide the poles of the gamma functions in the numerators. This means that the integrand is analytic on a strip centered around the imaginary axis. 
Now the integration over real time $t$ can be regularized by introducing cut-offs at $\pm T$ and then extended into Euclidean radial coordinate with cutoff near the Poincare boundary. Because all boundary terms vanish as we take away the cut offs it is safe to ignore the Euclidean contributions to the correlator and treat the real time computation in a distributional sense. The relevant integrals are of the form 
\begin{equation}
    \int_{-T}^T dt \; e^{i(k-2s)t}= 2\pi \;\delta_T\left(\frac{k}{2}-s\right),
\end{equation}
where $\delta_T(s)$ is a sine kernel peaked at $s=0$. As we take $T\rightarrow \infty$ the kernel approached a Dirac delta function. To compute the integration over $s$ we can deform the contour $\mathcal{I}$ to pass through the peak of the kernel since the integrand is analytic, then we can take the limit $T\rightarrow\infty$:
\begin{equation}
  \lim_{T\rightarrow \infty}  \frac{1}{i}\int_\mathcal{I} ds\frac{\Gamma\left(\frac{L}{2}-s\right) \Gamma\left(\frac{L}{2}+s\right) }{ \Gamma(L)} \;\delta_T\left(\frac{k}{2}-s\right)= \frac{1}{i}\frac{\Gamma\left(\frac{L-k}{2}\right) \Gamma\left(\frac{L+k}{2}\right) }{ \Gamma(L)}
\end{equation}What this procedure does is properly take into account the contribution of the delta function $\delta\left(\frac{k}{2}-s\right)$ that we would obtain from performing the $t$ integration in the most naive way. Putting everything together the one-point function is
\begin{equation}
\langle O_{L,k}\rangle= -2^{L+1}N\cos^4 \theta_0  \; \,\mathsf{N}_L\,\frac{\Gamma\left(\frac{L-k}{2}\right) \Gamma\left(\frac{L+k}{2}\right) }{ \Gamma(L)}\left[L D_L^{\theta} -\frac{1}{2} \sec^2\theta_0\frac{(L-k)(L+k)}{(L+1)}\right]\,\tilde{Y}_{(L,k)}(\theta_0, \phi_0).
\end{equation}
The integration over real time gives the same contribution as the integration over Euclidean time in previous calculations, except that the origin of the gamma function factors is more clear; they come from a product of wavefunctions of an scalar field in $AdS$. In the case of the $AdS$ giant graviton this contribution would be more interesting since the radial is non-trivial because the bulk propagator ends at a finite non-zero radius $r_0$.\\
The expressions can be simplified in terms of the variable $x=\cos(2\theta_0)$. In that variable the differential operator $L D_L^\theta$ takes a particularly nice form
\begin{equation}
    L D_L^\theta\left(\sin^k \theta_0 f(\cos2\theta)\right)= \left[\frac{1-x}{2}\right]^{\frac{k}{2}}\bigg((L-k)f(x)+ 2(1-x)\partial_x f(x)\bigg).
\end{equation}
The remaining expressions can be reduced using relations between Jacobi polynomials. For our purposes its better to keep everything in terms of the same type polynomials that appear in the spherical harmonics which gives
\begin{equation}
  \langle O_{L,k}\rangle=  \frac{\sqrt{L} \cos ^2(\theta ) \left(1+(-1)^{L-k}\right) i^{L-| k| +2}
   \sin ^{| k| }\theta_0 P_{\frac{ (L-| k| -2)}{2}}^{(| k| ,1)}(\cos 2
   \theta_0)}{L-| k| } e^{i k \phi_0}.
\end{equation}
This form of the correlator is not analytic as we take $|k|\rightarrow L$, even though there is a form of this correlator which is analytic \cite{Holguin:2022zii}.  These formulas are the same by relations of hypergeometric functions.

\subsection{Extremal Correlators}
In the case of extremal correlators the boundary terms that arise from integrating by parts do not vanish in as we go to the infinite past or future in Euclidean time, so we should be careful about solving the equations of motion, or equivalently quantizing the supergravity modes in the presence of the brane. Since the problematic term is non-zero both in Euclidean and Lorentzian signature we would try making sense of the variational problem in Lorentzian signature keeping in mind that we are implicitly placing a wavefunction for the brane at the infinite past and future. Since the problematic term is only supported at a point we expect that its renormalization does not affect the boundary terms that would be needed to make the variational problem well defined without the brane. This is more clear in the Euclidean picture where the coupling to the brane is localized at a point in the boundary and so its better to interpret it as a modification of the boundary conditions for the field $s_{\Delta,\pm\Delta}$. For maximally charged fields $s_{\Delta, \Delta}$, varying the action would lead to the condition
\begin{equation}
    \delta \left(\partial_t s_{\Delta,\Delta}-i\Delta s_{\Delta, \Delta}\right)\big|_{t \,= -T}=0.
\end{equation}
The other boundary contribution in this case can be safely rotated into Euclidean signature and its contribution vanishes at the boundary so we can drop it. The boundary condition is satisfied if 
\begin{equation}
    (\Delta+i\partial_t)s_{(\Delta,\Delta)}|_{t \,= -T} =\text{constant}=0
\end{equation}
The constant must vanish in order to preserve supersymmetry, or else the action is infinite in the Euclidean region. With a lot of hindsight we can recognize this as a BPS condition for the field $s_{L,L}$ which reflects the fact that the extremal correlators preserve the same supersymmetry algebra as a two point function of half-BPS states. The solution of the scalar equations of motion with the correct boundary conditions is then
\begin{equation}
    s_{\Delta,\Delta}= \mathsf{s}_{\Delta,\Delta}^\dagger \;e^{i \Delta t}
\end{equation}
which is to say that only the raising operator part of the field satisfies the boundary conditions. The phase factor $e^{i \Delta t}$ is what we would obtain from time evolving the operator insertion from $t=0_-$. This solution should correspond to an operator insertion at $t=-T$ in the bulk 
\begin{equation}
    \mathsf{s}_{\Delta,\Delta}^\dagger \ket{\mathcal{B}}.
\end{equation}
\begin{figure}
    \centering
    \includegraphics[width=0.45\linewidth]{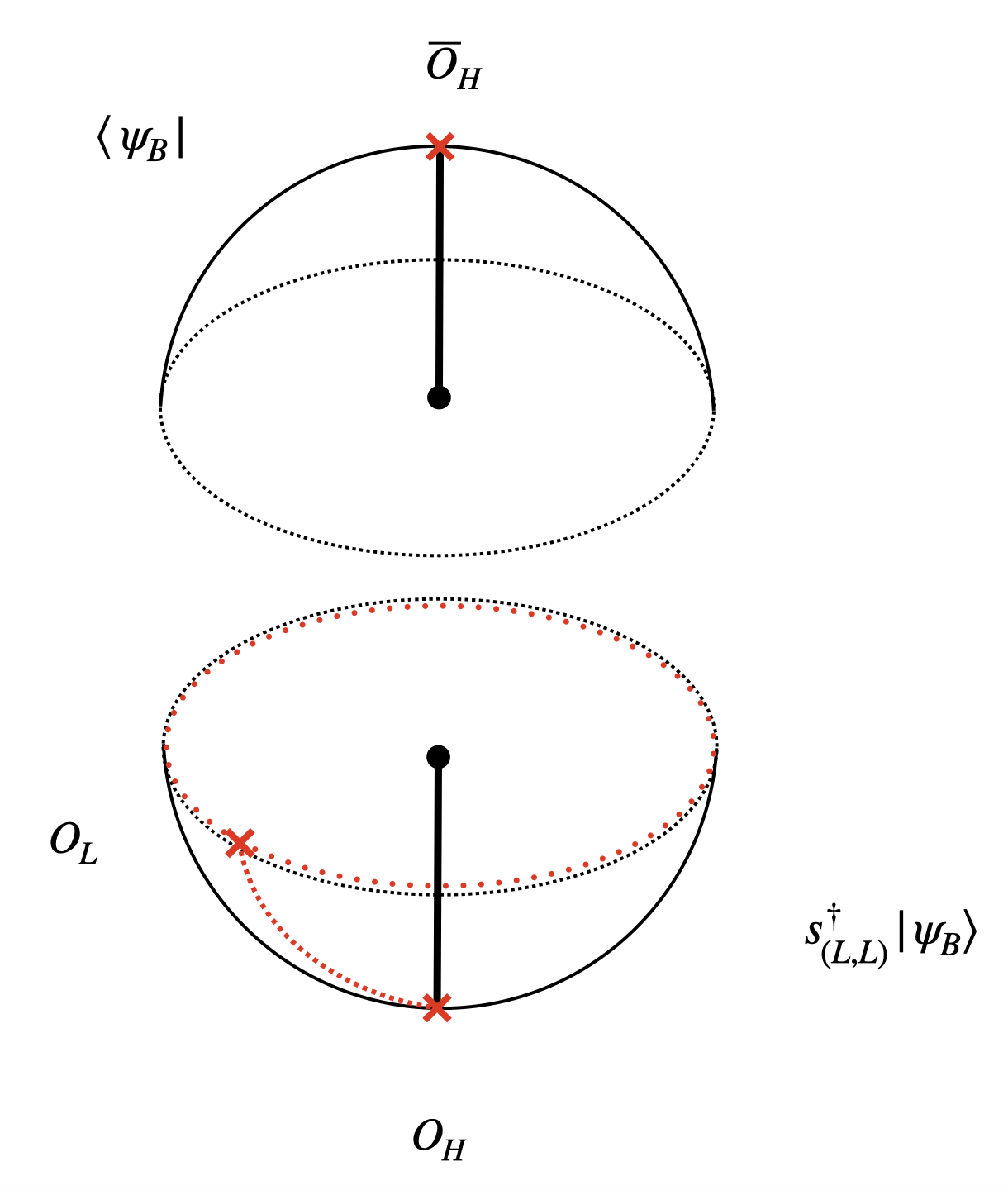}
    \caption{Sketch of the extremal three point function. The insertion of the light operator is taken to be on the Euclidean cap corresponding to the initial state. This inserts an additional half-BPS operator on the south pole of $S^4$ multiplied by a phase factor arising from the boundary propagator. }
    \label{fig:extremal}
\end{figure}
The state $\ket{\mathcal{B}}$ here denotes the wavefunction of the giant which is placed at $t=\pm T$. In string theory this boundary state would have a description in terms of half-BPS closed string fields, which can be approximated by $\mathsf{s}_{k,k}^\dagger$ in the large $N$ limit. Schematically we could write
\begin{equation}
   \ket{\mathcal{B}}\sim \exp\left(\sum_{k} t_k \,\mathsf{s}_{k,k}^\dagger\right)\ket{0}
\end{equation}
and so the extremal three point function would compute an overlap 
\begin{equation}
    \bra{\mathcal{B}} \mathsf{s}_{\Delta,\Delta}^\dagger \ket{\mathcal{B}} \sim t_\Delta.
\end{equation}
One interpretation of this formula is that the vevs of maximally charged operators directly measure the wavefunction of the boundary state $\ket{\mathcal{B}}$.
In general determining the form of the wavefunction $  \ket{\mathcal{B}}$ in supergravity would be a scheme-dependent task. For heavy backgrounds this is often done by solving the equations of motion for the metric of the new background and expanding near the boundary which leads to infinities that need to be regulated and renormalized appropriately. Most correlation functions, are insensitive to such renormalization, owing to the fact that they come from bulk effects. In the case of extremal correlators this is no longer true since changing renormalization schemes is equivalent to changing the wavefunctions of the initial and final states. For supersymmetric states there is often a supersymmetric renormalization scheme, but this choice cannot be unique either. For instance one could chose boundary conditions which allow or disallow certain multi-trace couplings of supersymmetric operators \cite{Witten:2001ua}. This could be interpreted as a change of basis for the boundary theory \cite{Arutyunov:2000ima}. Although this would not change most expectation values in the large $N$ limit, the values of vevs of maximally charged operators would change since one is explicitly changing the wavefunctions of the initial states. A way of resolving this scheme ambiguity is to input information from a UV completion of the supergravity effective action. One such completion is the boundary field theory, and this approach has been taken before in the renormalization of supersymmetric backgrounds. Another approach would be to compute the wavefunctions in the bulk string theory, while a more bottom-up approach would be insist that correlators are analytic in their charges. 

\section{Canonical Quantization and Brane Wavefunctions}\label{sec 4}
To compute the extremal correlators we will use results from the quantization of the half-BPS sector of type IIB supergravity. The presentations of these results mirrors closely the analysis of \cite{Bershadsky:1993cx, Dijkgraaf:2002fc, Aganagic:2003db, Aganagic:2003qj} in the context of topological strings; one particularly important aspect will be the role of a chiral boson in describing the deformation theory of the moduli space of supergravity solutions. To describe the wavefunctions of giant graviton branes we will need to be careful about flux quantization in the quantum theory. As expected from bosonization the only consistent choice is for the brane functions to be related to vertex operators of the chiral boson. 
\subsection{Quantizing LLM}
The most general half-BPS solution to type IIB supergravity was described by \cite{Lin:2004nb}. The metric can be put in the form
\begin{equation}
    \begin{aligned}
    &ds^2 = \mathcal{H}^{-2}\left( -(dt+V)^2+ \left(\frac{1}{2}-\zeta\right)d\Omega_3^2 + \left(\frac{1}{2}+\zeta\right)d\Tilde{\Omega}_3^2\right)+ \mathcal{H}^{2}\left(dy^2+ \delta_{ij}dx^idx^j \right)\\
    &d\left(\star_3 \frac{d\zeta(x_i,y)}{y}\right)=0,\;\;\;\;\;\;
    y dV= \star_3 d\zeta,\;\;\;\; \mathcal{H}^{-2}= \frac{y}{\sqrt{\frac{1}{4}-\zeta^2}}.
    \end{aligned}
\end{equation}
A particular geometry is determined from the boundary conditions for the function $\zeta(x_i,y)$ at $y=0$. For smooth geometries this function is fixed to be piecewise constant with values $\pm \frac{1}{2}$ on a 2d plane spanned by $x_i$, which separates the plane into droplet regions. 
The half-BPS moduli space of asymptotically $AdS_5\times S^5$ solutions can be quantized by considering the space of area preserving diffeomorphisms on the droplets \cite{Lin:2004nb,Grant:2005qc, Maoz:2005nk}.  The ADM Hamiltonian for these class of solutions is given by 
\begin{equation}\label{free fermions}
H= N^2\int_\mathcal{D} \frac{dx dp}{2\pi} \,\frac{x^2+p^2}{2}- E_0,
\end{equation}
where $E_0$ is the energy of the vacuum $AdS_5\times S^5$. The geometries are completely specified by a collection of droplets $\mathcal{D}$ in a two dimensional plane. The coordinates of this plane are identified with the collective coordinates of the gauge theory degrees of freedom. The Hamiltonian \eqref{free fermions} is equivalent to the Hamiltonian of $N$ free fermions in a harmonic oscillator potential. 
For our purposes it is enough to consider the case where $\mathcal{D}$ is a small deformation of the unit disk where the deformation is classically described by a chiral boson
\begin{equation}
(\partial_t- \partial_\phi)r^2(\phi)=0.
\end{equation}
The symplectic form  
\begin{equation}
    \omega= \frac{dx\wedge dp}{\pi}= \frac{1}{2\pi} dr^2\wedge d\phi,
\end{equation}
measures the area of the two dimensional plane.
From this one can define a one-form potential which is closely related to the chiral boson
\begin{equation}
  A=  pdx=  \frac{1}{2}r^2(\phi)d\phi.
\end{equation}
In the quantum theory $r^2(\phi)$ becomes an operator with the mode expansion:
\begin{equation}
    r^2(\phi)= \frac{1}{\pi}+ \sum_{n>0}\left(\alpha_n e^{i n \phi}+ \alpha_{-n} e^{-i n \phi}\right).
\end{equation}
The modes $\alpha_{-k}$ can be identified with metric perturbations that do not leave the half-BPS sector \cite{Grant:2005qc}. More generally we have a collection of $w_\infty$ generators corresponding to the area preserving diffeomorphism on the plane.
Often we will $r^2(\phi)$ as a function of the complex coordinate $z=  e^{i\phi}$  restricted to the boundary of the droplet. The resulting field $\chi(z)$ is also a chiral boson in the $(t, \phi)$ variables and should not be confused with a homorphic current. \\
\subsection{Brane Wavefunctions}
\begin{figure}
    \centering
    \includegraphics[width=0.6\linewidth]{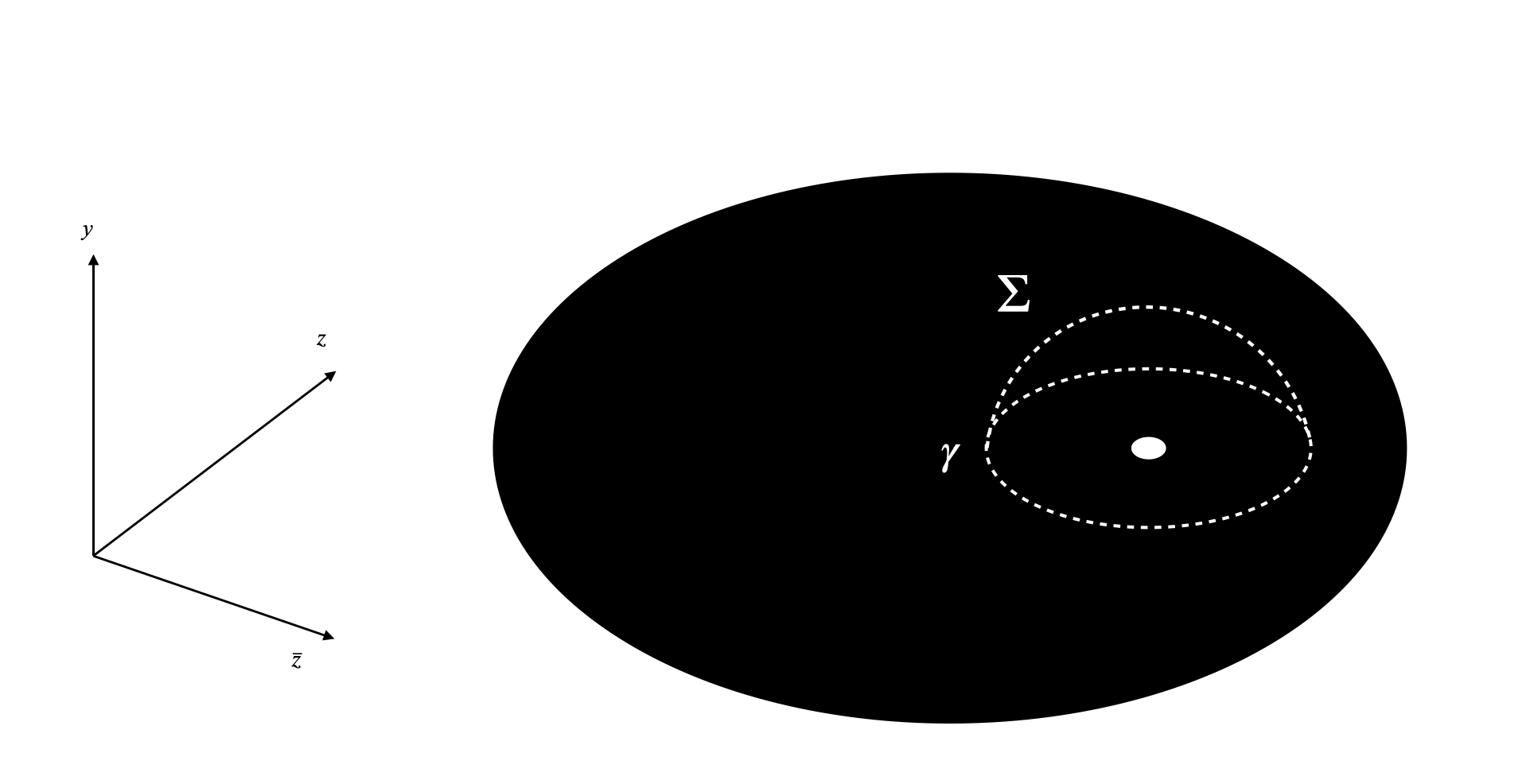}
    \caption{An example of a test surface $\Sigma$ over which we measure the five form flux. The surface extends along the $y$ direction of the geometry and an $S^3$ is fibered over it creating a non-trivial five-cycle. }
    \label{fig:LLMflux}
\end{figure}
In the presence of a brane the situation is modified since the one-form is no longer globally defined inside the droplet. This can be seen as follows. Consider a curve $\gamma$ which encloses a collection of $k$ branes. We can take a surface $\Sigma_2$ extended along the $y$ direction ending on $\gamma$ and integrate $F_5$ on $S^3\times \Sigma_2$. The value of this integral measures the number of sphere giant gravitons inside of $\gamma$. On the other hand using Stokes' theorem the integral reduces to an integration over a one-form along $\gamma$
\begin{equation}
    \frac{k}{N}=\int_{S^3\times \Sigma_2}F_5= \int_\gamma \hat{B}.
\end{equation}
The field strength associated to $\hat{B}$ at $y=0$ is related to the density of eigenvalues by the formula 
\begin{equation}
\begin{aligned}
 (\partial_p \hat{B}_x-\partial_x \hat{B}_p)dx\wedge dp&= -\frac{1}{4}y^3 \star_3 \partial_y\left(\frac{\zeta+1/2}{y^2}\right)dy= \frac{1}{2}(\zeta+1/2)dx\wedge dp-\frac{1}{4} y\partial_y \zeta dx\wedge dp\\
 \hat{F}_{px}(y=0)&= \frac{1}{2}(\zeta(y=0)+1/2)dx\wedge dp.
\end{aligned}
\end{equation}
where we used the fact that the regularity of the background implies that $\zeta(y)\simeq \pm \frac{1}{2}+ f(x,p) y^2 + O(y^4)$. Strictly speaking we have assumed that the metric is non-singular which is not quite true if one tries to compute the backreaction due to a small number of branes. Instead we will treat additional branes as probes. Within a black region $\hat{F}_{px}(y=0)$ naively vanishes, but the presence of the branes changes the value of the integration over $\hat{B}$. Since the contour can be pushed to be arbitrarily close to the branes we have
\begin{equation}
   \frac{k}{N}= \hat{n}_\gamma=\int_\gamma  \hat{B}= \oint_{z_0}pdx.
\end{equation}
If $k$ is of order one we can ignore the effect of backreaction, meaning that no additional boundaries are created, at the cost of introducing singularities in the symplectic potential. In the quantum theory we should should upgrade $p$ into an operator.  The operator $\hat{n}_\gamma$ measures the number of sphere giants contained withing the contour $\gamma$. For now we may set $k=1$. Now consider the operator $\mathcal{B}(z_0)$ which creates a sphere giant at some position $z_0=r_0 e^{i\phi_0}$. Within correlation functions we must have the identity
\begin{equation}
   \big\langle \dots \int_\gamma  \hat{B}\; \mathcal{B}(z_0)\dots \big\rangle= \big\langle \dots \oint_{z_0}p dx\; \mathcal{B}(z_0)\dots \big\rangle= \frac{1}{N} \big\langle \dots \mathcal{B}(z_0)\dots \big\rangle.
\end{equation}
This means that the wavefunction of the giant is given by 
\begin{equation}
    \psi_\mathcal{B}=e^{iN \int^{P} p\, dx}= e^{N \int^{P} \bar{z}\,dz}=e^{N \chi(z_0)}
\end{equation}
where $P$ denotes the position of the brane in the LLM plane. This is a consequence of the fact that the field $r^2(\phi)$ has the same OPE rules as a chiral boson. In the complex coordinates the field $\chi(z_0)$ would also correspond to a chiral boson but with a different radius $z_0= r_0 e^{i\phi}$\footnote{Chiral bosons play an analogous role in the Kodaira-Spencer theory describing deformations of local Calabi-Yau threefolds. }.  Because of the BPS condition the wavefunctions $\psi_\mathcal{B}$ are not globally defined functions on $S^4$, but transform as sections of a line bundle. This is related to the fact that the wavefunctions explicitly depend on the integration path to the position of the brane. The wavefunctions at the north and south pole are related to one another by canonical transformations of the half-BPS sector. This is the statement that the fields carrying a negative $R$-charge are identified as momentum variables of those with positive charge. The supersymmetry condition restricts the possible insertions on the north and south poles of $S^4$. To perform calculations we will need to move the wavefunctions to lie in a single hemisphere of $S^4$ so that we can use the OPE.

To check that this is consistent we can compute the norm of the state created by the wavefunction in the large $N$ limit:
\begin{equation}
   |\psi_\mathcal{B}|^2= \bigg\langle e^{N \int^{\Bar{z}_0} z\,d\Bar{z}} \;e^{N \int^{z_0} \Bar{z}\,dz}\bigg\rangle= e^{N (|z_0|^2-1)}= e^{-N \cos^2\theta_0}.
\end{equation}
This is done by bringing both wavefunctions to the south pole, putting the end point of the integration at the Fermi surface, and substituting the classical values for the chiral boson $\bar{z}$. Notice that although each wavefunction is separately not a globally defined function, the norm of the state is well defined since the expressions in the exponents combine into a closed form. The operator that creates the final state of the brane is related to the one that creates the initial state by a canonical transformation. For an AdS giant the position of the point $P$ lies outside the disk giving the following wavefunction
\begin{equation}
   |\tilde{\psi}_\mathcal{B}|^2=e^{N (|z_0|^2-1)}= e^{N \sinh^2\rho_0}.
\end{equation}
This agrees with the large $N$ value of the norm of the coherent state operators on the gauge theory, which gives some confidence to our proposal. A next check would be to compute the expectation values of maximally charged operators in the background of the brane.

In order to compute correlation functions one should upgrade the coordinates to operators. The approach of \cite{Grant:2005qc, Maoz:2005nk} was to treat the variable which is conjugate to the coordinate along the Fermi surface. Said coordinate generically satisfies the equations of motion of a chiral boson. Equivalently one could have with the field $p(x)$ or $\bar{z}(z)$. The expansion of $\Bar{z}(z)$ into modes is
\begin{equation}
\begin{aligned}
     \bar{z}(z)&= \partial\chi(z)= \alpha_0 +\sum_{n>0}\left(\alpha_{n}z^{n-1}+ \alpha_{-n} z^{-n-1}\right)\\
     [\alpha_{n}, \alpha_{m}]&= \frac{n}{N^2} \,\delta_{n+m,0}.
\end{aligned}
\end{equation}
This is familiar from the complex matrix model and the operators $\alpha_n$ correspond to classical couplings $t_n$. For large $|z|$ these correspond to non-normalizable modes of the area preserving diffeomorphisms, meaning that in the classical limit they should describe vevs of maximally charged operators. The expression for the brane wavefunction is of the form
\begin{equation}
    \ket{\psi_B}= \exp\left[-N \sum_{n>0} \frac{1}{n}\alpha_{-n} z_0^{-n}\right] \ket{0}
\end{equation}
which is reminiscent of the expression of the determinant operator as an exponetiation of single traces. The bulk operator dual to $\tr Z^k$ would be the closed string field $\alpha_{-k}$, and so acting on the brane state gives
\begin{equation}
:\psi_{\mathcal{B}}^\dagger\;\,\alpha_{-k} \,\psi_{\mathcal{B}}: \;\sim  \,-\frac{1}{N} z_0^k\,|\psi_{\mathcal{B}}|^2+ \dots,
\end{equation}
where the dots include terms which are suppressed in the large $N$ limit. After normalizing by the norm of the state created by $\alpha_{-k}$ gives the value for the extremal correlator
\begin{equation}
    \bigg \langle \psi_{\mathcal{B}} \bigg|\; \frac{N\alpha_{-k}}{ \sqrt{k}}\;\bigg|\psi_{\mathcal{B}}\bigg \rangle= -\frac{z_0^k}{\sqrt{k}}\times |\psi_{\mathcal{B}}|^2= -\frac{\sin^k\theta_0 \,e^{ik \phi_0}}{\sqrt{k}}\times |\psi_{\mathcal{B}}|^2.
\end{equation}
This is exactly the expectation value of $O_k= \frac{1}{\sqrt{k}}\tr Z^k$ in the background of a coherent state in the large $N$ limit, including the wavefunction factors and the minus sign. This should come at a surprise, given that in principle the wavefunction of the state can be renormalized from weak to string coupling and in general can be scheme dependent. More precisely what we showed is that the matrix elements within the half-BPS sector are not renormalized, and certainly matrix elements with more generic operators will get non-trivial corrections.  
\subsection{Some comments on LLM correlators}
Finally we make some comments on the case where the background operators describe a backreacted bubbling geometry \cite{Lin:2004nb} as opposed to a probe. Some of these correlators where studied holographically in \cite{Skenderis:2007yb} and more recently in \cite{Turton:2025svk, Turton:2025cnn}. In \cite{Anempodistov:2025maj} a class of complex matrix models were proposed for computing one point functions of chiral primaries in generic LLM backgrounds\footnote{For earlier works using free fermion methods see \cite{Takayama:2005yq}}. One important feature of these calculations is the necessity of normal ordering of the matrix model observables, much like in calculations for the gauge theory on $S^4$ \cite{Rodriguez-Gomez:2016cem}. This is explained by the fact that on a non-trivial background operator mixing with the identity operator is allowed, and the proper chiral primary operators on $S^4$ are combinations of those on $\mathbb{R}\times S^3$. This means that for general LLM backgrounds the expectation values of single trace chiral primaries operators on $\mathbb{R}\times S^3$ are computed by expectation values of \textit{multi-trace} observables on the matrix model, and it is these that should agree with the one point functions computed holographically in \cite{Skenderis:2007yb}. These non-linearities are expected from operator mixing \cite{Skenderis:2006uy}, but their connection to the mixing on $S^4$ is somewhat surprising.

The connection between LLM correlators and those of the theory on $S^4$ is clearer in the case of the extremal correlators, where one can relate the computation to one on $S^4$ via localization. This is because the insertion of additional chiral/anti-chiral operators can be interpreted as inserting additional operators on the north/south pole of the sphere. On the gravity side this calculation is closely related to to the analysis of \cite{Dijkgraaf:2002fc} in the context of the B-model. In fact the quantization of the LLM moduli space is essentially equivalent to a subsector of the B-model studied in \cite{Aganagic:2003qj}. This is what we find in the case of probe branes, where the extremal correlator comes from computing overlaps of closed string fields. What should then be the interpretation of non-extremal one-point functions? Unlike their extremal counterparts these cannot be related to operator insertions on the poles of the $S^4$, but rather to scalar insertions on the equator of the sphere. In the probe brane case we saw that there is a Euclidean calculation in which the operators are inserted at the equato and the calculation reduced to a Witten diagram calculation in the Euclidean geometry. When we insert order $N$ branes, the picture significantly changes and the Euclidean caps should be replaced by a backreacted geometry. The one-point functions are then read off from the asymptotic form of the fields away from the poles of the boundary $S^4$. For this reason the calculation of one-point functions in heavy backgrounds do not include any integration over the Euclidean time $\tau$ which could be interpreted as an orbit average, despite the fact that a general half-BPS background would break rotational and dilatation symmetries separately. We should also emphasize that the non-extremal single trace operators that we are dealing with are closely related to the generators of the $w_\infty$ algebra describing area preserving diffeomorphisms of the LLM droplet. In the context of the B-model, the Ward identities for the broken $w_\infty$ generators are powerful enough to determine all the closed string field correlators. It would be interesting to understand the role of these symmetries in the context of LLM correlators as they might shed some light on how to obtain the relatively simple matrix model expressions found in \cite{Anempodistov:2025maj} from the gravity description.

\section{Discussion}\label{discussion}
In this paper we revisited the computations of extremal and non-extremal one point functions of chiral primaries in the backgrounds of giant gravitons. We clarified the role of averaging and wavefunctions in the computation of one-point functions of charged chiral primaries, demystifying the apparent conflict between the previously obtained results in the literature. As a result we conclude that holographic computations can be performed with or without averaging and find perfect agreement with gauge theory resutls. This gives further confidence to proposals of \cite{Berenstein:2013md, Berenstein:2022srd, Holguin:2022zii} for the dual description of BPS coherent states in $\mathcal{N}=4$ SYM. We also clarified the computation of extremal correlators involving probe giants, by identifying an incorrect implementation of the holographic dictionary in the prescriptions of \cite{Zarembo:2010rr, Bissi:2011dc}, and proposed a bulk computation for extremal correlators that reproduces the large $N$ results for the two and three point functions of half-BPS operators. It is important to emphasize that although the results we find are not new, fully clarifying issues with semiclassical calculations in holography is essential for tacking more interesting issues, such as computations of three heavy operators. It is worth emphasizing that in order to compute the correct form of the brane wavefunctions in supergravity it was crucial to properly account for potential anomalies due to five-form flux quantization. This should be contrasted with the approach taken recently in \cite{Lee:2023iil, Caputa:2025ikn}, where the value of $N$ was analytically continued away from being an integer. The resulting evanescent operators can probably be understood from anomaly considerations. It would be interesting to try reproduce correlation functions involving trace relation operators with a bulk calculation. It should be expected that this a more covariant approach to quantizing the string fields, since properly quantizing the fluxes will eliminate any unphysical states. 

One natural direction to pursue is study correlation functions involving open and closed strings \cite{Bak:2011yy}, for instance $\frac{1}{4}-\frac{1}{2}-\frac{1}{2}$ BPS correlators which are protected for certain values of the operator charges \cite{Bissi:2021hjk}. On the gauge theory these involve additional modifications of determinant operators describing the ground state of open string ending on a giant. These correlators are important given the renewed interest in bootstrapping scattering of strings off branes in AdS \cite{Alday:2023mvu, Alday:2024srr, Chen:2025yxg, Chen:2025cod}. Similarly it would be important to further develop the semiclassical methods for computing correlation functions of heavy BPS operators. Particularly the correlation function of three giants is likely to shed light on the various proposals for computing three point functions of excited string states \cite{Zarembo:2010rr,Buchbinder:2010vw, Bajnok:2014sza, Janik:2011bd}.

\acknowledgments
I am happy to thank Pedro Vieira, Vladimir Kazakov, and Harish Murali for discussions that motivated this work. I also thank Robinson Mancilla for discussions regarding holographic renormalization. The author acknowledges conversations with ChatGPT used to simplify hypergeometric functions. The work of A.H. is supported by ERC-2022-CoG - FAIM 101088193. A.H. thanks the Perimeter Institute for Theoretical Physics where a part of this work has been done. 
\appendix
\section{Gauge Theory Semiclassics}
Here we review some of the necessary details of the gauge theory computation of the three point function involving semiclassical states. We will work in the language where two of the states create a half BPS background $\ket{\mathcal{B}}$, not necessarily having a fixed conformal dimension. The residual symmetry of this state is $PSU(2|2)^2$ without any additional central extension. This is a consequence of the fact that for a generic half-BPS background the $SO(2)_R$ charge symmetry is broken. Without loss of generality we consider inserting an additional half-BPS operator of fixed dimension at $t=0$. The symmetry fixes the form of the correlator to be
\begin{equation}
    \frac{\bra{\mathcal{B}} O_L(\tilde{n}, t=0)\ket{\mathcal{B}}}{\bra{\mathcal{B}}\ket{\mathcal{B}}}= \sum_{k=-L}^{L} (n\cdot \tilde{n})^{L/2} (\bar{n}\cdot \tilde{n})^{L/2} \times \left(\frac{n\cdot \tilde{n}}{\bar{n}\cdot \tilde{n}}\right)^{k}a^{\mathcal{B}}_{L,k}
\end{equation}
where the physical information is encoded in the one point function $a^{\mathcal{B}}_{L,k}$, which is the quantity that should be compared with the holographic calculation. The coefficient $a^{\mathcal{B}}_{L,k}$ is generically a sum of OPE coefficients, and the precise structure constants may be recovered if the full details of the wavefunctions of $\mathcal{B}$ are known. \\
The example we will have in mind is that the background state is created from only $Z$ fields. One choice of $O_L(\tilde{n})$ would be for example
\begin{equation}
    O_L(\tilde{n})= \frac{1}{\sqrt{L}}\tr\bigg[\left(\frac{Z+ \bar{Z}+ Y- \bar{Y}}{2}\right)\bigg].
\end{equation}
If we chose the $R$-symmetry polarizations such that they are unit normalized the first kinematic factor would $2^{-L}$ without loss of generality, while the second measures a phase which in this case is equal to one $e^{i\psi_0 }=1$. The one-point function would then be of the form
\begin{equation}
    \frac{\bra{\mathcal{B}} O_L(\tilde{n}, t=0)\ket{\mathcal{B}}}{\bra{\mathcal{B}}\ket{\mathcal{B}}}= \frac{1}{2^L}\sum_{k=-L}^{L}\,a^{\mathcal{B}}_{L,k}.
\end{equation}
Note that each term in the sum comes from a particular $SO(2)$ charge contribution. 
Generically the one-point function will be a function of the parameters of the background state and not just a number. 
First lets consider the case where the background state is
\begin{equation}
   \ket{ \mathcal{B}_{\text{sphere}}} = \det(a_z^\dagger-z_0)\ket{0}.
\end{equation}
The norm of the state in the saddle point approximation is
\begin{equation}
    \bra{\mathcal{B}_{\text{sphere}}}   \ket{ \mathcal{B}_{\text{sphere}}} = e^{N(|z_0|^2-1)}= e^{-N \cos^2\theta_0}
\end{equation}
and the stability of the saddle implies that $|z_0|= \cos\theta_0<1$. In that case the correlator becomes 
\begin{equation}
\begin{aligned}
     \bra{\mathcal{B}_{\text{sphere}}} O_L(\tilde{n}, t=0)\ket{\mathcal{B}_{\text{sphere}}}&= -\frac{1}{2^{L}\sqrt{L}\mathcal{Z}_\rho}\int d\rho\; e^{-N \rho \bar{\rho}} \;(|z_0|^2+ |\rho|^2)^N\; \tr \left[\begin{pmatrix}
         z_0& -i \rho\\
         -i \bar{\rho}& \bar{z}_0
     \end{pmatrix}^{-L}\right]\\
     &= \frac{1}{2^L}\times \left[-\, \frac{2}{\sqrt{L}}T_{L}\left(\frac{z_0+\bar{z}_0}{2}\right)\right]\times e^{N(|z_0|^2-1)}.
\end{aligned}
\end{equation}
The overall minus sign is there since the leading Wick contraction of the fermions requires us to move a single fermion the past $2L-1$ others in the trace. The polynomial $T_L(x)$  is a Chebyshev polynomial, and we can read off the contributions from each charged sector by gathering terms by their degree in $\frac{z_0}{\bar{z}_0}$, which is equivalent to expanding in Fourier modes of the phase of $z_0$. 

For example the maximally charged contribution is
\begin{equation}
a_{(L,L)}^{\mathcal{B}_{\text{sphere}}}= -\frac{|z_0|^L}{\sqrt{L}}=  -\frac{\sin^L \theta_0}{\sqrt{L}};
\end{equation}
which corresponds to an extremal three point 
function. 

\section{Global Coordinate Propagators}\label{propagators}
We start by considering a massive scalar field in global $AdS_5$. The metric in global coordinates is given by 
\begin{equation}
    ds^2= -(r^2+1) d\tau^2+ \frac{dr^2}{r^2+1}+ r^2 d\Omega_3^2,
\end{equation}
and the equations of motion for a scalar field are 
\begin{equation}
\begin{aligned}
    &\frac{1}{\sqrt{-g}}\partial_\nu \left(\sqrt{-g}g^{\mu \nu} \partial_\mu \phi\right)- m^2 \phi\\
    &=- \frac{1}{r^2+1 }\partial_\tau^2\phi + \frac{1}{r^{d-2}}\partial_r\left((r^{d-2}(r^2+1)\partial_r \phi\right)+ \frac{1}{r^2}\Delta_{S^{d-1}}\phi -m^2 \phi=0.
\end{aligned}
\end{equation}
The equation is separable and has solutions of the form
\begin{equation}
    \phi(r,\tau,\Omega)= \sum_{n,k} \left(\phi_{n,k }e^{i \omega_{n,k} \tau}+\phi^*_{n,k }e^{-i \omega_{n,k} \tau}\right) R_{n,k}(r) Y_k(\Omega).
\end{equation}
The solution for the radial equation is a sum of hypergeometric functions. Imposing that the solution is regular at the origin and at infinity fixes the form of the radial modes and gives a quantization condition for $\omega$:
\begin{equation}
\begin{aligned}
    R_{n,k}(r)&=\mathsf{N}_{n,k}\, r^k \left(r^2+1\right)^{\frac{1}{2} (\Delta +k+2 n)} \,
   _2F_1\left(\frac{d}{2}+k+n,k+n+\Delta ;\frac{d}{2}+k;-r^2\right) \\
   \omega_{n,k}&= \pm\left(\Delta+k + 2n\right).
\end{aligned}
\end{equation}
The normalization constants are determined by evaluating the Klein-Gordon inner product of the solution mode by mode. They are 
\begin{equation}
    \mathsf{N}_{n,k}^{-1}= (-1)^k \sqrt{\frac{n!\, \Gamma^2
   \left(\frac{d}{2}+k\right)
   \Gamma
   \left(\frac{2-d}{2}+n+\Delta
   \right)}{\Gamma
   \left(\frac{d}{2}+k+n\right)
   \Gamma (k+n+\Delta )}}.
\end{equation}
The bulk-to-bulk propagator can be obtained by quantizing $\phi$ and computing its two point correlator. We will need the bulk-to-boundary propagator. For a bulk point at $r=0$ this is particularly simple and we can resum the series

\begin{equation}
 \sum_{n=0}^\infty \frac{(\Delta-1+n)\Gamma(2-\Delta)}{n!\,\Gamma(2-n-\Delta)}  e^{i(t-t')(2n+\Delta) }= (\Delta-1) \frac{2^{-\Delta}}{\cos^\Delta(t)}.
\end{equation}
The sum can be simplifies using reflection formulas for the gamma function 
\begin{equation}
 K_\Delta(t, r; 0)=   \sum_{n=0}^\infty (-1)^n  \frac{\Gamma (n+\Delta )
   e^{i t (\Delta +2 n)}}{n!
   \Gamma (\Delta )}.
\end{equation}
One useful rewriting of the sum is to express it as a sum over residues via a Mellin integral formula. 
\begin{equation}
  K_\Delta(t, r; 0)= \frac{1}{2\pi i}\int_{-i \infty-\frac{\Delta}{2}+\epsilon}^{i \infty-\frac{\Delta}{2}+\epsilon} ds   \; \frac{\Gamma\left(\frac{\Delta}{2}-s\right) \Gamma\left(\frac{\Delta}{2}+s\right) }{ \Gamma(\Delta)} e^{-2it s}.
\end{equation}
The sign is chosen by closing the contour to the right. This expression has the advantage of being essentially symmetric in $s\leftrightarrow -s$. The gamma functions have poles at $s= -\frac{\Delta}{2}, -\frac{\Delta}{2}-1,-\frac{\Delta}{2}-2, \dots $ and $s= \frac{\Delta}{2}, \frac{\Delta}{2}+1,\frac{\Delta}{2}+2, \dots$, so we may deform the contour to pass through the imaginary axis. 


%\begin{thebibliography}{99}

% Please avoid comments such as "For a review'', "For some examples",
% "and references therein" or move them in the text. In general,
% please leave only references in the bibliography and move all
% accessory text in footnotes.

% Also, please have only one work for each \bibitem.

%\end{thebibliography}

\bibliographystyle{JHEP}
	\cleardoublepage

\renewcommand*{\bibname}{References}

\bibliography{references}
\end{document}